\documentclass[useAMS,usenatbib]{mn2e}
\usepackage{graphicx}

\usepackage{subfigure}
\usepackage{booktabs}
\usepackage{verbatim}
\usepackage{lscape}
\usepackage{threeparttable}

\title[Spectropolarimetry of Herbig Ae/Be stars]{Linear spectropolarimetry across the optical spectrum of Herbig Ae/Be stars\thanks{Based on observations collected at the European Southern
    Observatory (ESO), Paranal, Chile under programme 088.C-0858(A), 088.C-0858(B) and 088.C-0858(C).}}
\author[]{}

\author[K. M. Ababakr, R. D. Oudmaijer and J.S. Vink]{K. M. Ababakr$^{1}$\thanks{E-mail:
pykma@leeds.ac.uk}, R. D. Oudmaijer$^{1}$ and J.S. Vink$^{2}$\\
  $^{1}$School of Physics and Astronomy, University of Leeds, EC Stoner Building, Leeds LS2 9JT, UK\\
$^{2}$Armagh Observatory, College Hill,  Armagh, UK\\}

\begin{document}

\date{Accepted 2016 June 22. Received 2016 June 22; in original form 2016 April 29}

\pagerange{\pageref{firstpage}--\pageref{lastpage}} \pubyear{2002}

\maketitle

\label{firstpage}

\begin{abstract} 

We present the results of spectropolarimetric observations of 12
Herbig Ae/Be objects. Our data have the largest spectropolarimetric
wavelength coverage, 4560~$\rm\AA$ to $9480~{\rm \AA}$, published to
date. A change in linear polarisation across the H$\alpha$ line, is detected in all
objects. Such a line effect reveals the fact that stellar photons are
scattered off free electrons that are not distributed in a spherically
symmetric volume, suggesting the presence of small disks around these accreting
objects. Thanks to the large wavelength coverage, we can report that
H$\alpha$ is the spectral line in the optical wavelength range that is
most sensitive to revealing deviations from spherical symmetry, and
the one most likely to show a line effect across the polarisation in
such cases. Few other spectral lines display changes in polarisation
across the line. In addition, H$\alpha$ is the only line which shows
an effect across its absorption component  in some
sources. We present a scenario explaining this finding and demonstrate
that the detection of the line effect strongly relies on the number of
photons scattered into our line of sight. We highlight the special
case of R Mon, which is the only object in our sample to show many
lines with a polarisation effect, which is much stronger than in all
other objects. Given that the object and its nebulosity is spatially resolved, we argue that this is due to scattering of the stellar
and emission spectrum off circumstellar dust.

\end{abstract}


\begin{table*}
 \centering
\caption{HAeBe Observations. The $\it V$magnitude and spectral type are listed in column 3 \& 4. The spectral type is taken from \citet{ababakr15} and \citet{Fairlamb2015} while the $\it V$ magnitude is taken from \citet{2003AJ....126.2971V} and SIMBAD. The integration times (column 8, 9 \& 10) denote the total exposures in ${\it V}$ (4560 - 5860~${\rm \AA}$), ${\it R}$ (5750 - 7310~${\rm \AA}$) and ${\it Z}$ (7730 - 9480~${\rm \AA}$)  bands.}

\begin{tabular}{l l l l l l l l l l }
\hline
Name      & Other name&  {\it V}&Spec. type& \multicolumn{3}{c}{Date}  &   \multicolumn{3}{c}{Exposure (s)} \\
           &    &    &    &  {\it V}  & {\it R}  & {\it Z}&  {\it V}  & {\it R}  & {\it Z}  \\
\hline 
PDS 27  &     DW CMa   &     13.0   &    B3       & 04-02-12     &    04-02-12     &   04-02-12&8$\times$340&24$\times$50&16$\times$120        \\
PDS 37  &     Hen 3-373   &     13.5   &    B3       & 05-02-12     &    05-02-12     &   05-02-12&16$\times$340&24$\times$70&16$\times$100        \\
CPD-485215  &     Hen 3-847   &     10.6   &    B6       & 20-01-12     &    10-03-12     &   03-03-12&8$\times$150&32$\times$15&16$\times$90        \\
R Mon  &     MWC 151   &    11.8   &    B8       & 19-02-12     &    01-02-12     &   01-02-12&4$\times$340&12$\times$85&12$\times$80        \\
  &        &      &          &    &    12-10-11     &   &&12$\times$85&        \\
V380 Ori  &      &     10.7   &    A0       & 03-01-12     &   03-01-12     &   10-11-11&8$\times$60&16$\times$20&12$\times$20        \\
  &      &       &           &     &   12-10-11     &  &&12$\times$20&        \\
PDS 133  &   SPH 6     &     13.1   &    B6       &     &    21-01-12     &   21-01-12&&16$\times$150&8$\times$340        \\
BF Ori  &        &     9.7   &    A2       &    &    12-10-11     &  12-11-11  &&8$\times$75&8$\times$90        \\
MWC 275  &     HD 163296   &     6.9   &    A1       &      &    01-04-12     &   30-03-12&&32$\times$2.5&16$\times$7.5        \\
GU CMa  &    HD 52721   &     6.6   &    B1       &      &    07-01-12     &  07-01-12&&24$\times$2&24$\times$5        \\
HD 104237  &   PDS 61   &     6.6   &    A7       &      &    16-01-12     &  16-01-12&&40$\times$1&24$\times$5        \\
HD 85567  &       &     8.6   &    B7       &      &    30-12-11     &  08-12-11&&8$\times$5&16$\times$30        \\
  &       &        &           &      &    30-03-12     &  &&8$\times$5&        \\
HD 98922  &    Hen 3-644   &     6.8   &    A0       &      &    30-12-11     &  16-01-12&&40$\times$1.5,16$\times$1&24$\times$5        \\
\hline
\label{obser}
\end{tabular}
\end{table*}

\section{Introduction}

Despite the important role of massive stars in the evolution of the
interstellar medium and unlike their counterpart low mass stars that
are thought to be formed via magnetospheric accretion
\citep{1998ApJ...492..743M}, their accretion mechanism is still open
to debate. Observationally, the detection of massive stars is
challenging as they reside far away and are deeply embedded in their
natal clouds. As an alternative to massive stars, the intermediate
mass Herbig Ae/Be (HAeBe) stars are the best candidates to address
this issue. They are optically visible and bridge the gap between low
and high mass stars. They were first identified by Herbig as having a
spectral type A or B with emission lines
\citep{1960ApJS....4..337H}. The current view is that HAeBe
stars are surrounded by disks through which material continues to
accrete onto the star (e.g. \citealt{ilee14}), although the
precise scenario is not known (for reviews on the topic, see \citealt{grady15, kraus15,beltran16}).

In order to make progress, one needs to study the circumstellar
environment through which the  material accumulates onto the star
via accretion channels. This requires methods capable of probing the
matter very close to the star, as this is where the accretion action happens.
A key point is the observation whether the ionised region around the
star is spherically symmetric at small scales or not. If it is not,
then the possibility that a flattened structure is observed lends
support to the disk accretion scenario responsible for the formation
of such stars. Studying the circumstellar environment at these small
scales is possible through linear spectropolarimetry, measuring the
scattering of photons off free electrons in a dense, ionised gas.

The idea of spectropolarimetry is that in the ionised region, free
electrons scatter and polarise continuum photons from the central
star. If the geometry is not circular on the sky, for example in the
case of flattened disk, a net polarisation can be detected. However,
emission photons undergo less scattering as they emerge further away
from the central star. If we are to measure the polarisation as a
function of wavelength, this difference in scattering will be visible
as a change in polarisation across the line, which is often simply
referred to as the ``line effect''. The observed de-polarisation
across the H$\alpha$ line constituted the first evidence that the class
of Be stars were surrounded by disks \citep{1974MNRAS.167P..27C,
  1976ApJ...206..182P}.  This was only much later confirmed by direct
observations (e.g. \citealt{quirrenbach94} based on image
reconstruction of interferometric data; \citealt{wheelwright12}
using sub-milliarcsecond precision spectro-astrometry).

\citet{1999MNRAS.305..166O}, \citet{Vink2002} and \citet{Mottram07} extended
the use of the technique to HAeBe objects. \citet{Vink2002} found that
7 out of 12 Herbig Be (HBe) objects they observed have a
depolarisation line effect across H$\alpha$, very similar to what was
found in Be stars. As the line-effect becomes less pronounced for
lower inclinations, and would disappear for a face-on disk, which is
circular on the sky, the detection statistic strongly suggests that
all HBe stars are surrounded by small disks with sizes of order
several stellar radii.
In contrast, they found a different line effect for Herbig Ae (HAe)
objects, where enhanced {\it polarisation} across the H$\alpha$ was
found in 9 out of 11 stars. They proposed that the line itself is
intrinsically polarised since part of the emission lines originate
from a compact region, where the accretion takes
place. \citet{vink05a} found that HAe stars have a similar
spectropolarimetric signature as the lower mass T Tauri stars with
HBe stars having a different signature.

\citet{McLean1979} reported a different line effect, across the
absorption component of the emission line which is often called the
McLean effect.  The general idea is that the absorption blocks the
unscattered light from the beam, and photons originally emitted in
different directions are scattered into the line of sight, resulting
in enhanced polarisation across the absorption. An alternate
  hypothesis was provided by \citet{kuhn07}, who proposed the
  polarisation can be caused by selective absorption due to optical
  pumping of an anisotropic radiation field. In addition,
  \citet{kuhn11} point out that resonant line scattering, which potentially
  also produces line polarisation, predicts that the lines within the
  {Ca \sc ii} near-infrared triplet around 8500~$\rm \AA$ will be
  differently polarised from each other. For a recent review on the
  use of linear spectropolarimetry, see \citet{vink15}. 

In this work we aim to expand the existing spectropolarimetric work
that was mostly aimed at H$\alpha$ by observing other emission lines
probing different volumes and conditions than H$\alpha$. Here we
present a spectropolarimetric study of a sample of 12 HAeBe
objects. The new feature of the current study is the broad wavelength
range from 4560~$\rm\AA$ to $9480~{\rm \AA}$, covering almost the
entire optical spectrum.  The paper is organised as follows. In
Section 2 we discuss the sample selection criteria, the details of the
observations and data reduction. In Section 3 we present the results
starting with continuum polarisation and then discussing line
spectropolarimetry. The analysis is provided in Section 4. We
conclude in Section 5.

\section{Observations}
\subsection{Source selection}

Twelve objects were selected for this work from the X-shooter project
of 91 HAeBe stars \citep{Oudmaijer11, Fairlamb2015}. The main reason
for choosing these, was  the presence of strong emission lines
which allows for a good determination of their spectropolarimetric
properties. The targets were originally selected from the HAeBe
catalogue of \citet{1994A&AS..104..315T} and candidates of
\citet{2003AJ....126.2971V}. They are presented in Table \ref{obser},
alongside their spectral type and {\it V} magnitudes. The spectral
type is taken from \citet{ababakr15} and \citet{Fairlamb2015} while
the {\it V} magnitude is taken from \citet{2003AJ....126.2971V} and
SIMBAD. The results of PDS 27 and PDS 37 have been published elsewhere
in a study dedicated to these two objects, but we keep them in here
\citep{ababakr15}.

\subsection{Spectropolarimetric observations}

The spectropolarimetric data were obtained with the FORS2 spectrograph
mounted on ESO's VLT in Chile. The 1400V, 1200R and 1027Z gratings
centred at 5200~$\rm\AA$, 6500~$\rm\AA$ and 8600~$\rm\AA$ respectively
were used with a $4096\times2048$ pixel CCD to cover the entire
optical range from 4560~$\rm\AA$ to $9480~{\rm \AA}$. A spectral
resolution of R = 3800, 4800, 4800 was achieved in the {\it V}, {\it R},
and {\it Z} bands respectively, using the 0.5 arcsec slit. The dates and
the exposure times for  each band are given in Table \ref{obser}.

The optical polarimeter in FORS2 was employed for the linear
polarisation observations. It consists of a rotating half-wave plate
and a calcite block. The half-wave plate can be set to various angles
to get {\it QU} Stokes parameters while the calcite block separates
the light into two perpendicularly polarised light beams, the ordinary
(O) and extraordinary (E) beam. Four exposures with the half-wave
plate set at angles $0^{\circ}$, $45^{\circ}$, $22.5^{\circ}$, and
$67.5^{\circ}$ provide a complete set of observations. At each angle,
four spectra are obtained; the O and E beam of the object and sky
respectively. Several cycles of observation at the above 4 angles were
taken for each object to check for the consistency of the polarisation
and its features. Several polarised standard stars and unpolarised
standard stars were observed to calibrate the instrumental
polarisation and angle offset.

Spectropolarimetry is a photon-hungry method. Typical line effects are
of order 0.5\% and to properly sample the polarisation behaviour as a
function of wavelength, a polarisation precision of order 0.1\% per
pixel is therefore required. This corresponds to signal-to-noise
ratios of order 1000. We used long exposure times for faint objects
and several short exposure times for objects with strong H$\alpha$
lines in order to avoid saturation of the CCD.

\begin{figure*}
        \centering
        \begin{subfigure}{}
                \centering
                \includegraphics[width=18cm]{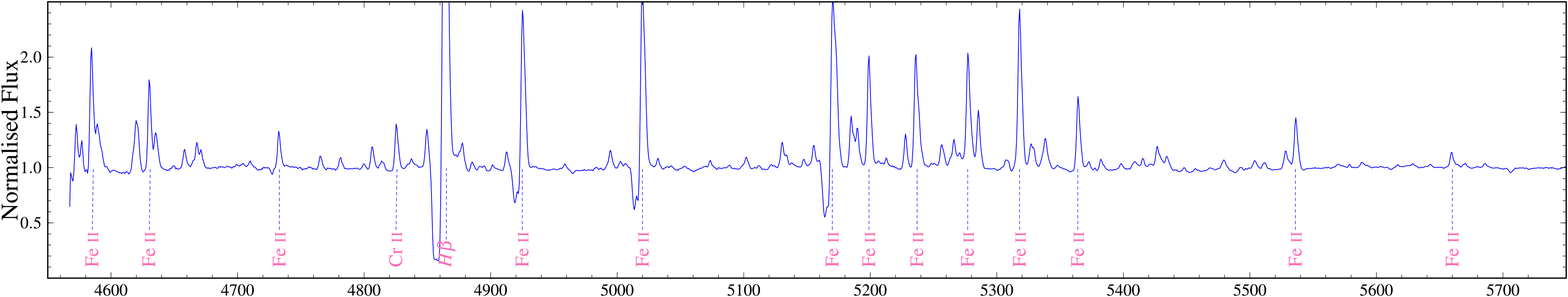}
                
        \end{subfigure} \vspace{-13pt}
        \begin{subfigure}{}
                \centering
                \includegraphics[width=18cm]{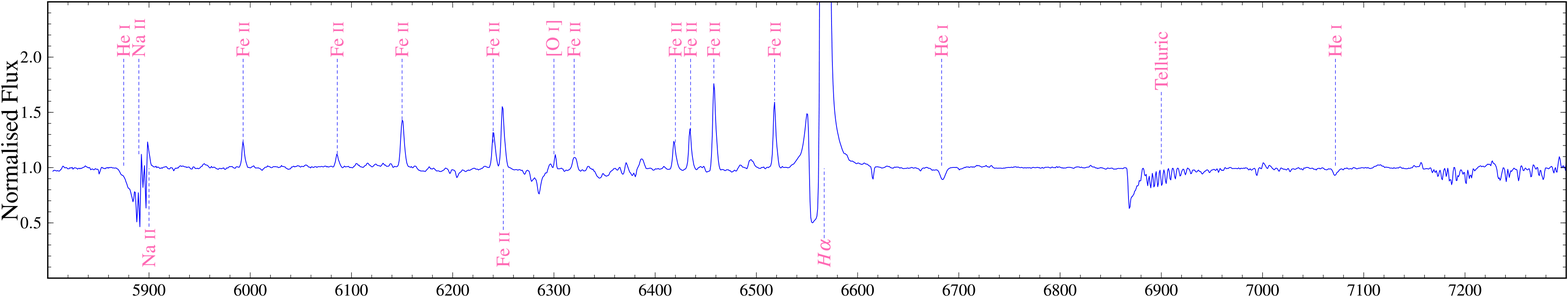}
                
        \end{subfigure}\vspace{-13pt}
       \begin{subfigure}{}
                \centering
                \includegraphics[width=18cm]{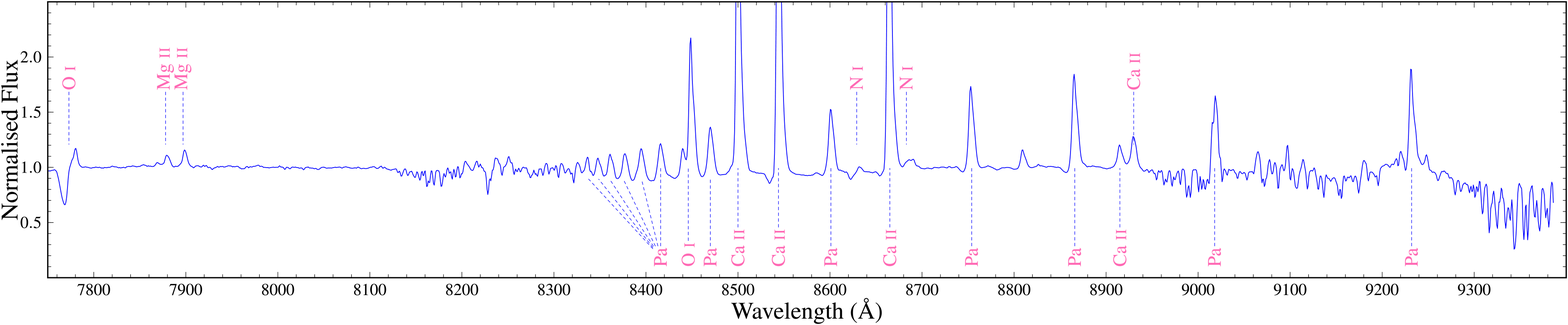}
                
        \end{subfigure}\vspace{-5pt}
        
        \caption{Normalised flux versus wavelength ($\rm \AA$) spectra of PDS 27 in the range from 4560~$\rm\AA$ to $9480~{\rm \AA}$, the observed spectral lines are identified.}
        \label{all27}
\end{figure*}


\subsection{Data reduction}

The data reduction was carried out using {\sc gasgano} and FORS
pipeline v4.9.18, which includes bias subtraction, flat fielding, sky
subtraction, extraction and wavelength calibration of the O and E
spectra. The Stokes parameters were determined from the O and E
spectra. Finally, the percentage of linear polarisation, P, and
polarisation angle, $\theta$, were obtained according to the following
equations:


\begin{equation}
 P = \sqrt[]{Q^{2}+U^{2}}
\end{equation}  

\begin{equation}
\theta = \frac{1}{2} \arctan \bigg(\frac{U}{Q}\bigg)    
\end{equation}


For the analysis purposes, the data were imported into the {\sc iraf}
\citep{Tody93} and {\sc polmap} package \citep{Harries96} maintained
by Starlink.  The individual data sets were combined to improve the
quality of the data. Four polarised and three zero polarised standard
stars were selected from the sample of \citet{Fossati07} for
polarimetric calibrations. The instrumental polarisation is found to
be $\sim$0.16\%, from the observation of unpolarised standard
stars. The angle offset is found to be $0.5^{\circ}$ from the
observation of polarised standard stars. In order to clearly visualise
the line effect, the change in polarisation across the lines, we rebin
the polarisation spectra to a maximum polarisation error per bin. We
performed the rebinning of the spectra very carefully to avoid losing
the line effect signature. We did not correct the data for the
instrumental and interstellar polarisation. These two parameters only
add a, small, wavelength independent vector to the observed Q and U
parameters whereas our main aim in this work is the detection of
polarimetric change across the lines. However, as the polarisation is
a vector quantity, the interstellar polarisation could change the
appearance of the polarisation signature across the line (see
  e.g. \citealt{Oudmaijer1998} in HD 87643). Therefore, careful analysis
is needed to characterise the line effect.


\begin{figure*}
        \centering
        \includegraphics[width=18cm]{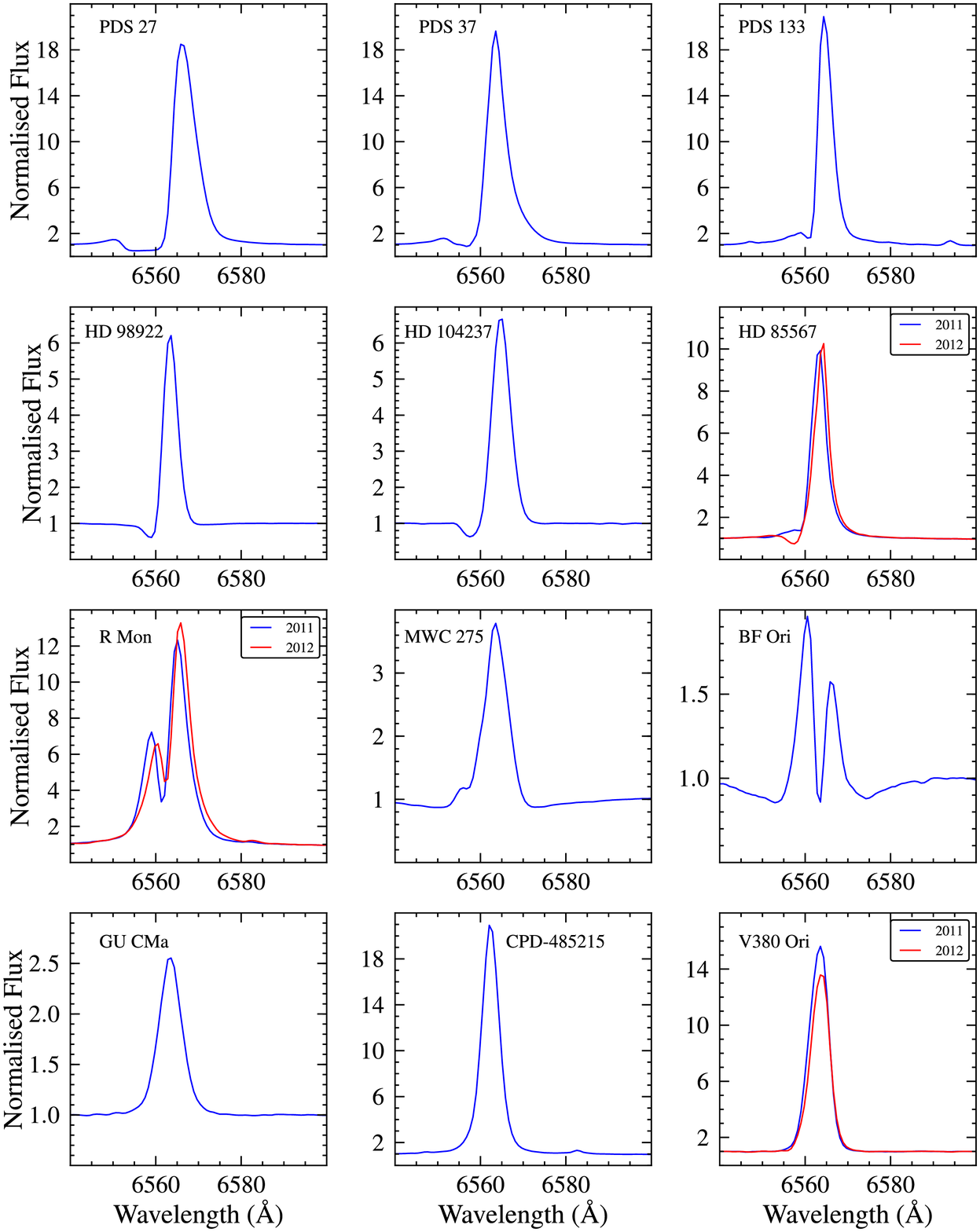}
        
   \caption{The H$\alpha$ line profiles, the spectra are
     normalised to one. The spectra of HD 85567, R Mon and V380 Ori
     are presented for two observation epochs: 2011 (blue solid line)
     and 2012 (blue solid line).}
     \label{alpha}
\end{figure*}

\subsection{The Optical Spectrum}

In the optical range from 4560~$\rm\AA$ to $9480~{\rm \AA}$, the
spectra show numerous spectral lines in emission and absorption.
The entire spectrum of
PDS 27 is presented in Fig. \ref{all27} to show the range of the
observed spectra and the identified spectral lines.

The H$\alpha$ lines are the strongest with line peak to continuum ratios
ranging from $\sim$ 2 to $\sim$ 20. In all targets, H$\alpha$ is in
emission, as shown in Fig. \ref{alpha}. The H$\alpha$ lines exhibit
various different profiles. These include pure emission lines, double
peaked emission with central absorption through to P Cygni
profiles. GU CMa, CPD-485215 and V380 Ori display a pure emission
line, R Mon and BF Ori show a double peaked emission line with central
absorption.
The remaining seven objects display P Cygni profiles, for two of which we
also have H$\beta$ data.
For both spectra that contain H$\beta$, the absorption component in
H$\beta$ is deeper than in H$\alpha$. In addition to Balmer
series, weak Paschen lines are also seen in emission for most objects.

The {Ca \sc ii} triplet lines at 8498, 8542 and 8662 $\rm \AA$ are
observed in all objects apart from GU CMa and CPD-485215. They are all
in emission apart from BF Ori where they are in absorption. The line
to continuum ratio is in the range 1.5 to 10.5. The {Ca \sc ii}
doublet at 8912 and 8927 $\rm \AA$ is present in all
objects apart from GU CMa, MWC 275 and CPD-485215. The lines are in
emission in all objects apart from HD 104237 and BF Ori where the
lines are in absorption.

Three {He \sc i} lines at 5876, 6678 and 7065 ~$\rm \AA$ are observed
in all objects. The lines are in absorption in six objects including
PDS 27, PDS 37, R Mon, HD 85567, GU CMa and BF Ori.  PDS 133, HD
104237, V380 Ori and HD 98922 have the lines in emission.
Inverse P Cygni profiles, weak emission and strong absorption {He \sc
  i} lines are detected in the spectra of MWC 275 and CPD-485215 and
indicate infalling material.

Several forbidden {[Fe \sc ii]} at 4814, 5158, 5262, 5333, 7155 and
8616 $\rm \AA$ lines are observed in the spectra of R Mon and
CPD-485215. {[S \sc ii]} at 6716 and 6730 $\rm \AA$ is also seen in
the spectra of R Mon. In addition, strong {[O \sc i]} at 6300, 6364
~$\rm \AA$ is present in the spectra of R Mon, CPD-485215 and PDS
133. However, {[O \sc i]} lines are either absent or very weak in the
spectra of the remainder of the targets. Unlike the cases of R
  Mon and CPD-485215, the {[O \sc i]} lines in the spectra of PDS 133
  are broad and double peaked which may suggest that the forbidden
  lines emerge from a rotating disk. This has been suggested before
  for HAeBe stars \citep{acke05}, although the broad line could also
  indicate a disk-wind (see e.g. \citealt{ignace06}).

In summary, we observe many strong emission lines which allows us to
investigate spectropolarimetric line effects which had never been done
before studied for most of these lines.

\begin{table*}
 \centering

\caption{The continuum polarisation of the targets measured in the
  following wavelength regions: {\it B} band centred at 4700~$\rm
  \AA$; {\it V} band centred at 5500~$\rm \AA$; {\it R} band centred
  at 7000~$\rm \AA$; {\it I} band centred at 9000~$\rm \AA$. The
  polarisation was measured over a wavelength range of 250~$\rm \AA$
  either side of the central wavelength, except for the {\it B} band
  where it was 125~$\rm \AA$ either side. All the errors are differential, where the instrumental error is $\sim$0.16\% in polarisation and $\sim$
$0.5^{\circ}$ in angle (see text in Section 2.3 for details).}
\begin{tabular}{l@{\hskip 0.1in} r@{\hskip 0.1in} r@{\hskip 0.1in} r@{\hskip 0.1in} r@{\hskip 0.1in} r@{\hskip 0.1in} r@{\hskip 0.1in} r@{\hskip 0.1in} r@{\hskip 0.1in}}
\hline
Object&\multicolumn{2}{c}{{\it B}}&\multicolumn{2}{c}{{\it V}}&\multicolumn{2}{c}{{\it R}}&\multicolumn{2}{c}{{\it I}}\\
& $P_{cont}(\%)$&$\theta_{cont}^{\circ}$&$P_{cont}(\%)$&$\theta_{cont}^{\circ}$&$P_{cont}(\%)$&$\theta_{cont}^{\circ}$&$P_{cont}(\%)$&$\theta_{cont}^{\circ}$\\
\hline 
PDS 27   &  8.60~$\pm$~0.03  &  18.60~$\pm$~0.10   &   9.00~$\pm$~0.01  &   18.90~$\pm$~0.01   &   8.80~$\pm$~0.01  &   18.40~$\pm$~0.04   &  7.50~$\pm$~0.01  &  18.10~$\pm$~0.03 \\
PDS 37   &  4.70~$\pm$~0.04  &  130.00~$\pm$~0.30  &   5.10~$\pm$~0.01  &   130.80~$\pm$~0.01  &   5.20~$\pm$~0.01  &   130.00~$\pm$~0.06  &  4.40~$\pm$~0.01  &  133.40~$\pm$~0.05 \\
CPD-485215   &  1.25~$\pm$~0.01  &  19.30~$\pm$~0.30  &   1.30~$\pm$~0.01  &   15.30~$\pm$~0.06  &   1.70~$\pm$~0.01  &   9.50~$\pm$~0.20  &  1.90~$\pm$~0.01  &  6.70~$\pm$~0.20 \\
R Mon   &  11.20~$\pm$~0.04  &  80.60~$\pm$~0.10  &   11.20~$\pm$~0.02  &   80.00~$\pm$~0.05  &   12.15~$\pm$~0.01  &   67.70~$\pm$~0.04  &  12.10~$\pm$~0.01  &  60.15~$\pm$~0.03 \\
 &    &   &   &   &   8.10~$\pm$~0.02  &   69.40~$\pm$~0.07  &    &   \\
V380 Ori   &  0.60~$\pm$~0.01  &  90.80~$\pm$~0.70  &   0.70~$\pm$~0.01  &   93.20~$\pm$~0.30  &   0.70~$\pm$~0.01  &   98.40~$\pm$~0.06  &  0.10~$\pm$~0.01  &  124.00~$\pm$~3.00 \\
V380 Ori   &    &   &     &    &   0.70~$\pm$~0.02  &   95.15~$\pm$~1.00  &   &   \\
PDS 133   &    &    &    &     &   2.30~$\pm$~0.04   &   37.70~$\pm$~0.50  &  2.10~$\pm$~0.03  &  38.30~$\pm$~0.05 \\
BF Ori   &   &    &    &    &   0.70~$\pm$~0.01  &   44.00~$\pm$~0.06  &    &  \\
MWC 275   &   &    &    &   &  0.70~$\pm$~0.01   &    1.60~$\pm$~0.20 &   0.67~$\pm$~0.01 &  14.84~$\pm$~0.28 \\
GU CMa   &   &    &    &    &   1.60~$\pm$~0.01  &   14.60~$\pm$~0.01  &  1.40~$\pm$~0.01  &  14.45~$\pm$~0.09 \\
HD 104237  &   &    &    &    &   0.35~$\pm$~0.01  &   174.60~$\pm$~0.60  &  0.58~$\pm$~0.01  & 182.89~$\pm$~0.35  \\
HD 85567   &   &    &    &    &   0.38~$\pm$~0.01  &   133.90~$\pm$~0.60  &  0.23~$\pm$~0.01  &  154.02~$\pm$~0.61 \\
   &   &    &    &    &   0.26~$\pm$~0.02  &   134.50~$\pm$~2.10  &    &   \\
HD 98922   &   &    &    &    &   0.40~$\pm$~0.01  &   174.90~$\pm$~0.30  &  0.58~$\pm$~0.01 &  177.90~$\pm$~0.30 \\
\hline
\end{tabular}
\label{con}
\end{table*}

\section{Results}

\begin{table}
\centering

\caption{Previous measured continuum polarisation. Column 2 gives the band at which the polarisation was measured. Columns 3 and 4 list the measured polarisation percentage and polarisation angle. The references are presented in the last column.}

\begin{tabular}{l l r r l}
\hline
Object&Band&$P_{cont}(\%)$&$\theta_{cont}^{\circ}$ &Ref.\\
\hline 
  
PDS 37  &\it V&3.253~$\pm$~0.104&120.1~$\pm$~1.0&1\\ 
CPD-485215&\it V&0.348~$\pm$~0.035&20.7~$\pm$3.0~&1\\  
R Mon   &\it V&12.0~$\pm$~1.2&91~$\pm$~3&2\\
        &\it U&11.1~$\pm$~0.7&85.7~$\pm$~1.8&3\\
        &\it B&11.1~$\pm$~0.4&86.6~$\pm$~1.1&3\\
        &\it V&12.3~$\pm$~0.5&89.4~$\pm$~1.2&3\\
        &\it R&12.6~$\pm$~0.3&83.5~$\pm$~0.8&3\\
        &\it I&12.1~$\pm$~0.4&94.5~$\pm$~0.9&3\\
V380 Ori &\it V&0.98~$\pm$~0.10&86~$\pm$~3&2\\
        &\it R&1.26~$\pm$~0.01&96~$\pm$~1&4\\ 
        &\it U&0.08~$\pm$~0.16&151~$\pm$~57&3\\ 
        &\it B&0.80~$\pm$~0.06&88.8~$\pm$~2.1&3\\
        &\it V&0.98~$\pm$~0.06&86.2~$\pm$~1.8&3\\
        &\it R&1.31~$\pm$~0.14&74.4~$\pm$~3.2&3\\
        &\it I&1.25~$\pm$~0.18&84.9~$\pm$~4.2&3\\
BF Ori  &\it R&0.6~$\pm$~0.1&58~$\pm$~1&5\\
        &\it R&0.886~$\pm$~0.017&52.0~$\pm$~0.6&6\\
        &\it V&0.46~$\pm$~0.01&&7\\    
GU CMa  &\it R&1.15~$\pm$~0.01&19~$\pm$~1&4\\
        &\it R&0.8~$\pm$~0.1&24~$\pm$~1&5\\
        &\it R&1.726~$\pm$~0.006&27.0~$\pm$~0.1&6\\
HD 104237&\it V&0.032~$\pm$~0.063&167.2~$\pm$56.5~&1\\  
HD 85567 &\it V&0.478~$\pm$~0.035&105.7~$\pm$~2.0&1\\  
HD 98922 &\it U&0.515~$\pm$~0.087&152.8~$\pm$~5.0&8\\
         &\it B&0.201~$\pm$~0.070&168.4~$\pm$~14.1&8\\
         &\it V&0.235~$\pm$~0.051&160.5~$\pm$~6.0&8\\  
MWC 275 &\it V&0.417~$\pm$~0.035&36.0~$\pm$~2.5&1\\
        &\it V&0.225~$\pm$~0.012&29.8~$\pm$~1.6&8\\
        &\it R&0.284~$\pm$~0.014&35.5~$\pm$~1.5&8\\
        &\it I&0.351~$\pm$~0.018&38.3~$\pm$~1.5&8\\
        &\it V&0.02~$\pm$~0.01&&7\\
      
\hline
\end{tabular}

\begin{tablenotes}
      \small
      \item[] References. 1:\citet{2009ApJ...698.2031R}, the errors on the polarisation angle for this reference are calculated from 0.5 $\times$ $\arctan$ ({\it $\sigma$p}/{\it P}); 2:\citet{1992ApJ...397..613H}; 3:\citet{1979ApJ...227..185V}; 4:\citet{1999MNRAS.305..166O}; 5:\citet{2011A&A...532A..28W}; 6:\citet{vink05a}; 7:\citet{2001A&A...379..564O}; 8:\citet{1998A&AS..131..401Y}, we calculated the average values for MWC 275, taken on the same day. 

\end{tablenotes}
\label{con1}
\end{table}

\begin{figure*}
        \centering
        \includegraphics[]{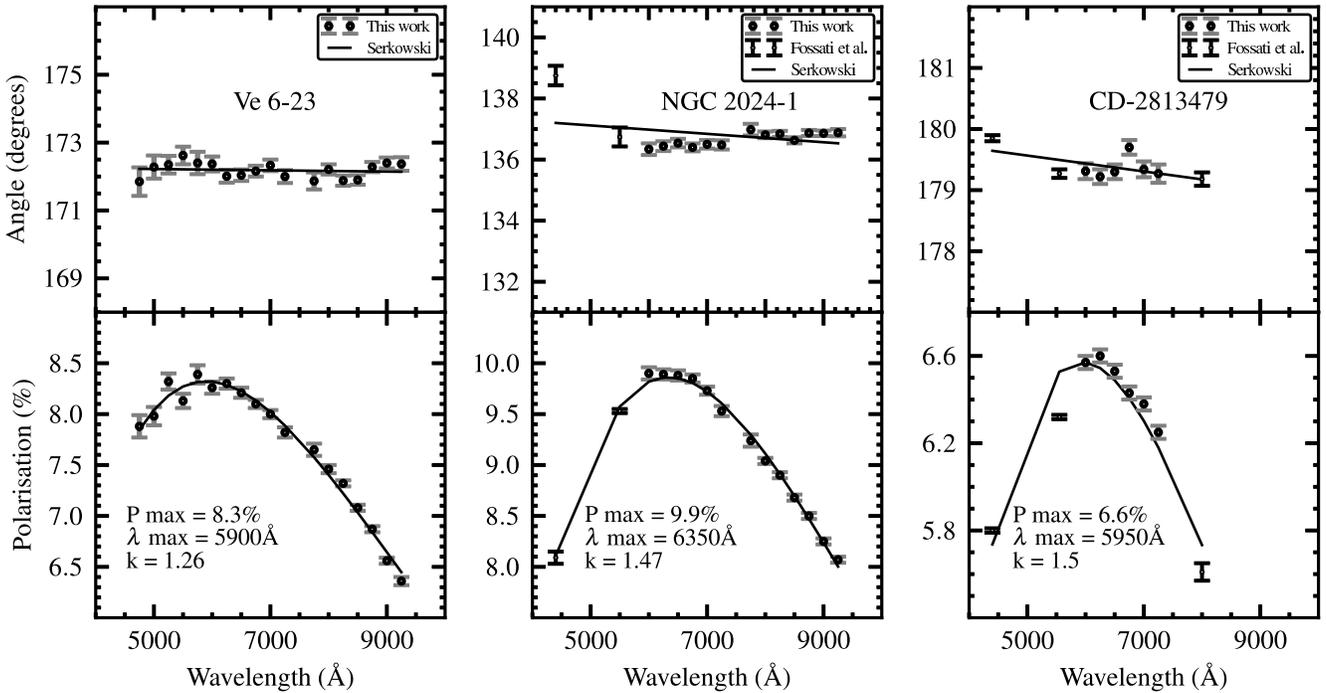}
        
   \caption{The polarisation and polarisation angles as a function of
     wavelength for the polarised standard stars. The data for this
     work are presented in black solid circles and were coarsely
     sampled to improve the error bars. The data in {\it B} and {\it
       V} for NGC 2024-1, and in {\it B}, {\it V} and {\it I} (black
     dots) are taken from \citet{Fossati07}. The black lines denote
     the best-fitting Serkowski law to the data. }
     \label{ve}
\end{figure*}

\begin{figure*}
        \centering
        \begin{subfigure}{}
                \centering
                \includegraphics[width=\textwidth]{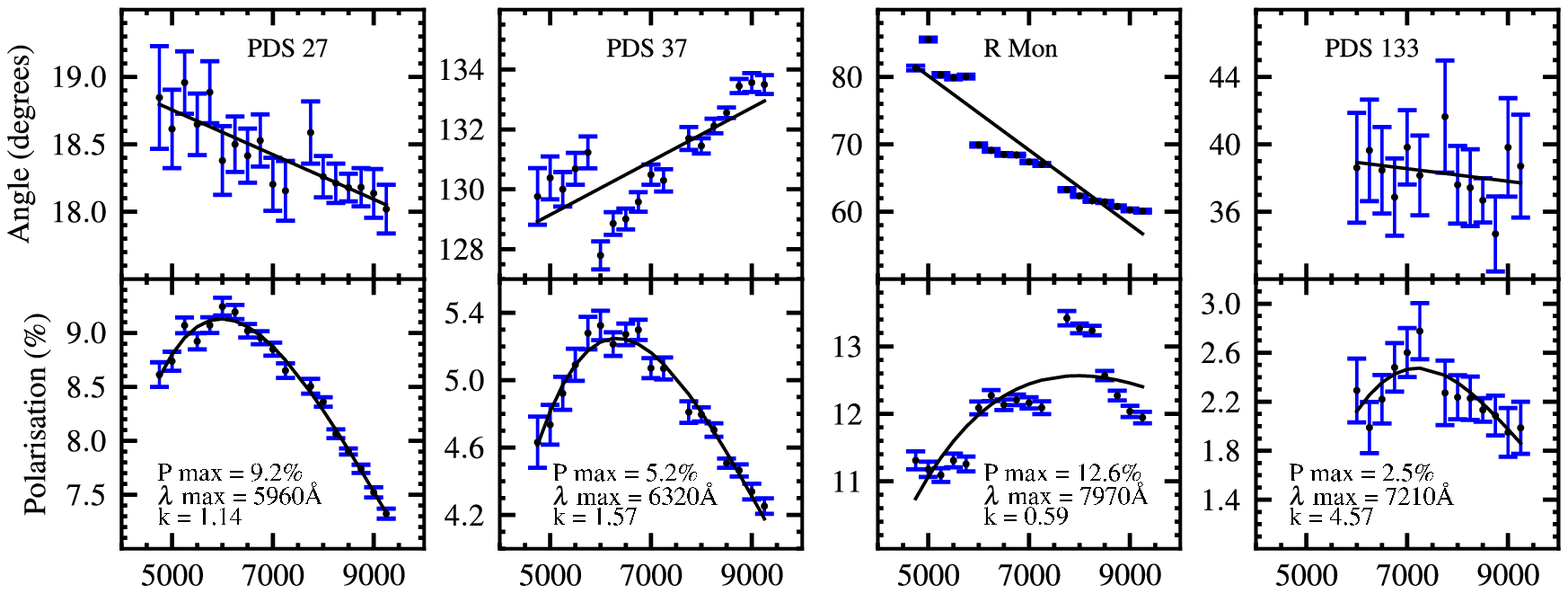}
                
        \end{subfigure}\vspace{-20pt}
        \begin{subfigure}{}
                \centering
                \includegraphics[width=\textwidth]{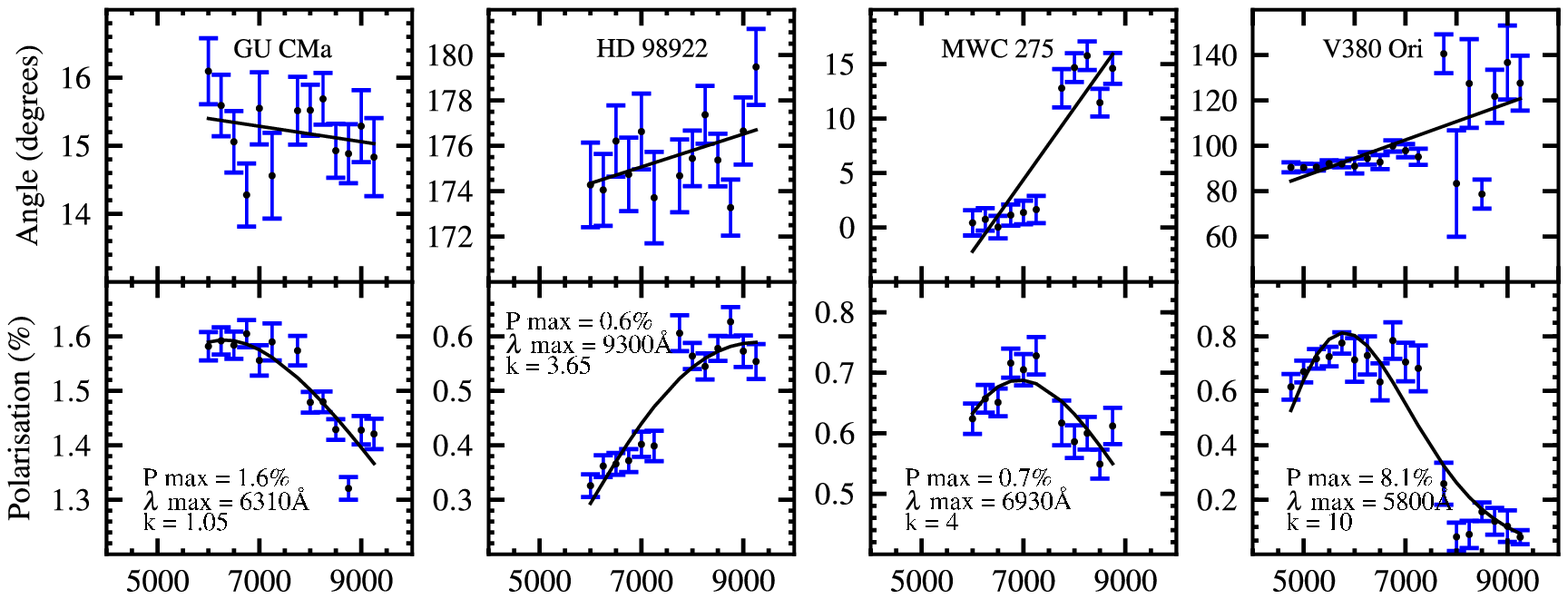}
                
        \end{subfigure}\vspace{-20pt}
       \begin{subfigure}{}
                \centering
                \includegraphics[width=\textwidth]{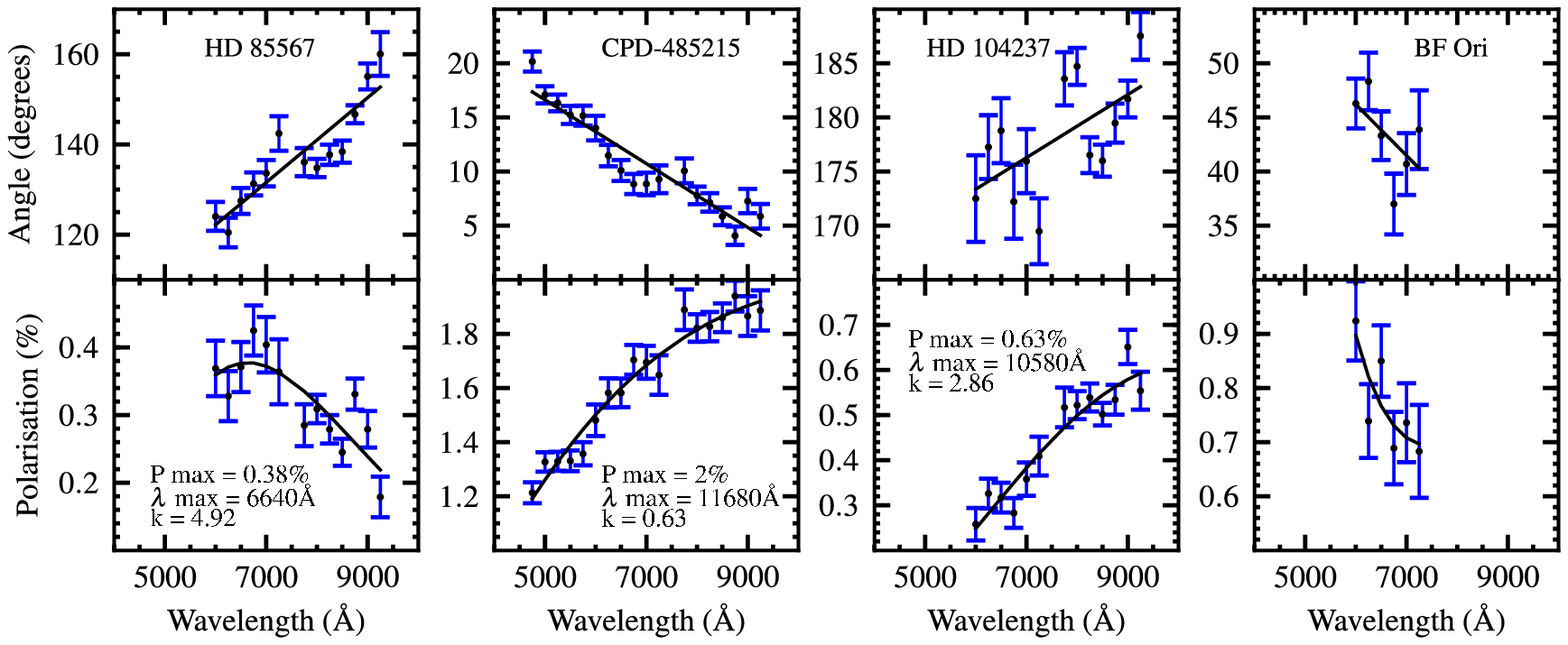}
                
        \end{subfigure}\vspace{-5pt}
        
        \caption{As the previous figure, but now for the HAeBe
          stars. The polarisation is shown in the lower panel while
          the polarisation angle is shown in the upper panel.  The
          black lines denote the best fitting Serkowski law to the
          data.}
        \label{allcon}
\end{figure*}  

\subsection{Continuum Polarisation}

The continuum polarisation is measured for the entire spectrum and the
results for the polarimetric standard stars are shown in
Fig. \ref{ve}, while those for the HAeBe stars are plotted in
Fig. \ref{allcon}. The data are rebinned using a coarse sampling of
$\sim$ 250 $\rm \AA$ per bin, to minimise errors.

Typically, the observed continuum polarisation is a vector addition of
the interstellar polarisation and the intrinsic polarisation which is
due to scattering of star light by circumstellar dust and free
electrons. The interstellar polarisation is caused by dichroic
absorption of aligned dust grains. Its wavelength dependence is well
understood and explained by \citet{1975ApJ...196..261S}. They found
that the linear polarisation follows an empirical curve according to
the following equation:

\begin{equation}
 \frac {P(\lambda)} {P_{max}} = {\rm exp}\bigg[-k~{\rm ln}^{2} \bigg(\frac{\lambda_{max}}{\lambda}\bigg)\bigg]
\end{equation}
where $\lambda_{max}$ is the wavelength where the polarisation is at
its maximum value, $P_{max}$, and {\it k} is the width of the
empirical curve. Typically, the polarisation peaks in the visual band
at $\sim$ 5500~$\rm\AA$, but can be in the range of
4500-8000~$\rm\AA$. {\it k} was initially taken to be a constant value
of {\it k} = 1.15. However, it has been found that the value of {\it
  k} depends on the value of $\lambda_{max}$, and is sensitive to the
size distribution of the dust particles \citep{Whittet92}.  In order
to assess whether the observed polarisation may have an interstellar
dust component, we fit the Serkowski law (Equation 3) to our targets
and polarised standard stars. $\lambda_{max}$, $P_{max}$ and {\it k}
are taken as a free parameters. The Serkowski law provides - as
expected - an excellent fit to the polarised standard stars as shown
in Fig. \ref{ve}. The figure also shows that the polarisation angle
remains constant over different bands.

The best fit to our targets is provided in Fig. \ref{allcon}. The
stars generally follow the Serkowski law but not as well as the
standard stars. In some objects, the polarisation spectra peak at much
longer wavelengths than the {\it V} band while, in addition, the
polarisation angle changes over different bands in all objects. This
overall behaviour would suggest that a fraction of the observed
polarisation is due to interstellar dust, but that an intrinsic
contribution is present. This is also consistent with the polarisation angle. For all objects,
the observed polarisation angle changes over different bands,
indicating the presence of intrinsic polarisation. The deviation of
the angle ranges from $1^\circ$ to $30^\circ$ (see
Fig.\ref{allcon}). The polarised standards did not show a
significant deviation in the observed polarisation angles.

We note that the continuum polarisation and the polarisation angle jump from one setting to another for some objects (see
Fig.\ref{allcon}). This is particularly striking in PDS 37, R Mon and MWC
  275. This might be because the objects
  are extended, and that a slightly different positioning of the slit
  could result in slightly different polarisation to be measured, which we shall discuss later. 

To facilitate comparison with past and future observations, we
measured the continuum polarisation at the {\it B}, {\it V}, {\it R}
and {\it I} bands and the results are summarised in Table \ref{con}.
The linear continuum polarisation ranges from $\sim$0.3 \% in the very
low polarised star HD 85567 to $\sim$12 \% in the highly polarised
star R Mon. The data of two epochs are provided for three objects, R
Mon, V380 Ori and HD 85567. R Mon shows a strong variability of order
of $\sim$4 \% in the {\it R} band between these two epochs. V380 Ori
shows similar polarisation values and HD 85567 shows a little
variation. The literature values for the objects are
summarised in Table \ref{con1}.  The polarised standard stars are
consistent with the literature values. The HAeBe stars are broadly
consistent with the previous measurements, but there is some variation
in the polarisation data.

The polarisation variability in the degree of polarisation and/or
polarisation angle is a common phenomenon among HAeBe objects
\citep{1994ASPC...62...63G,Jain95}. Its origin is thought to be caused
by the intrinsic polarisation over time as the interstellar
polarisation is unlikely to be variable over a short period of time.

\subsection{Line spectropolarimetry}

Most of the previous spectropolarimetric studies have been performed
on the strong H$\alpha$ emission lines. However, the advantage of this
study is that the long spectral coverage allows us to perform linear
spectropolarimetry on lines that had never been studied in this manner
before. We begin by discussing the results of the strongest hydrogen
recombination lines before we discuss other lines.


 \begin{figure*}
        
        \centering
       \includegraphics[width=16cm]{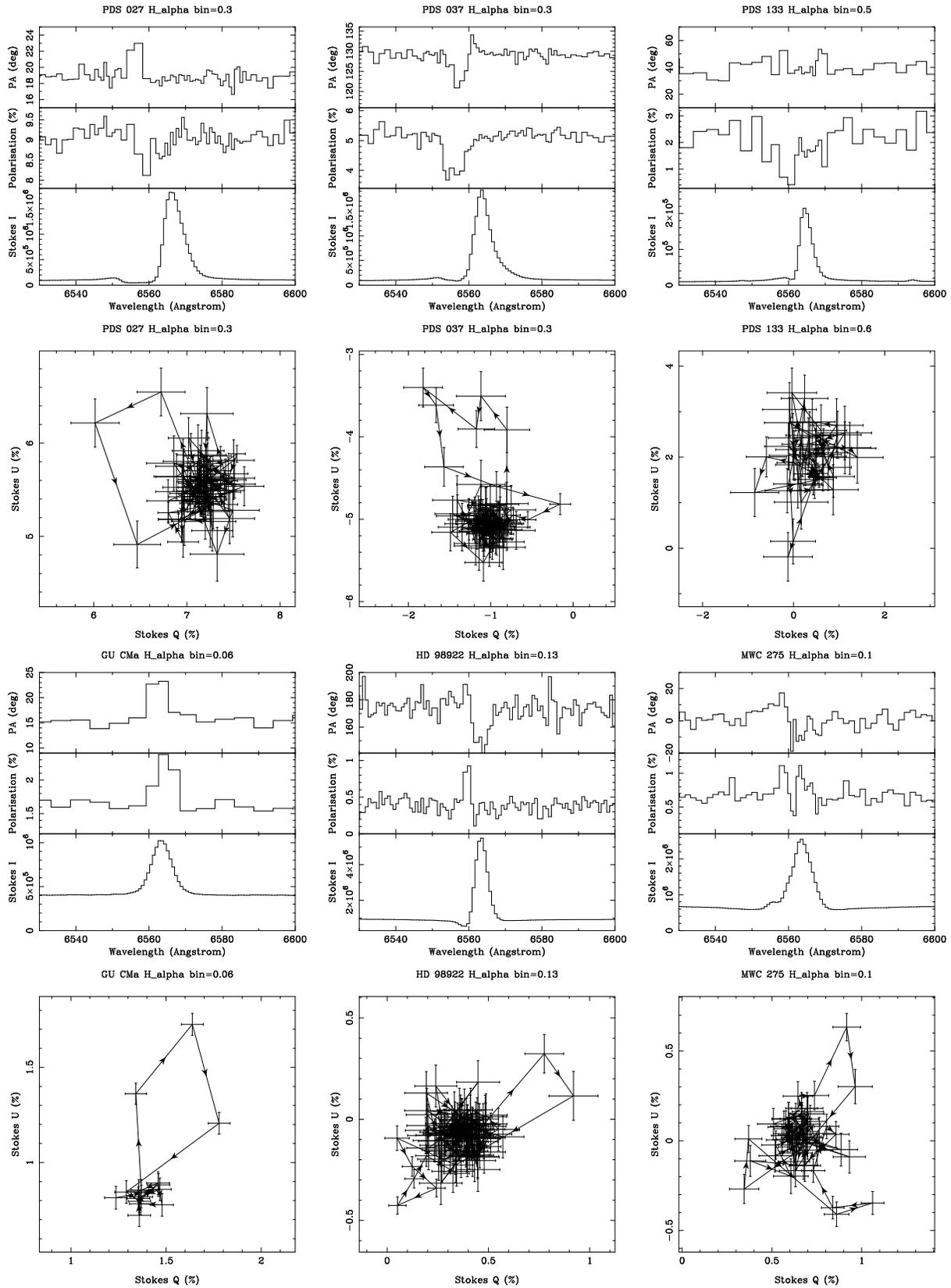}
   \caption{H$\alpha$ spectropolarimetry of the stars. The data are
     presented as a combination of triplots (top) and {(\it Q, U)} diagrams
     (bottom). In the triplot polarisation spectra, the Stokes
     intensity (I) is shown in the bottom panel, polarisation (\%) in
     the centre, while the position angle (PA) is shown in the upper
     panel. The Q and U Stokes parameters are plotted against each
     other below each triplot. The data are rebinned to a constant
     error in polarisation, which is indicated at the top of each
     plot. The arrows in the {(\it Q, U)} diagrams indicate the polarisation moves in and out of the line effect from blue to red wavelengths.}
    \label{halphafig}

\end{figure*}

  \begin{figure*}
\setcounter{figure}{4} 

\centering
        \includegraphics[width=16cm]{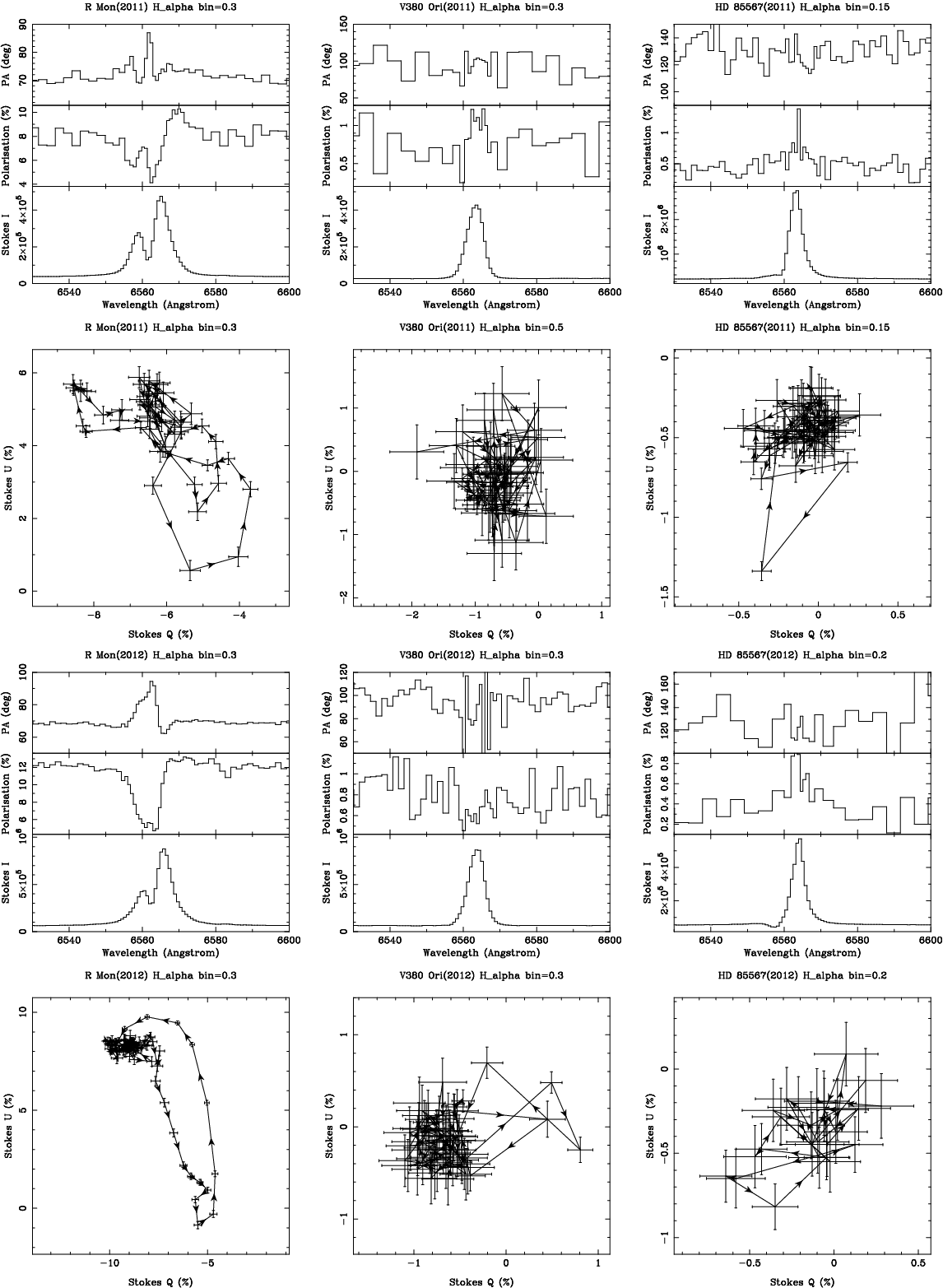}
           \caption{continued}
   \end{figure*}

 \begin{figure*}
\setcounter{figure}{4} 
        
        \centering
       \includegraphics[width=16cm]{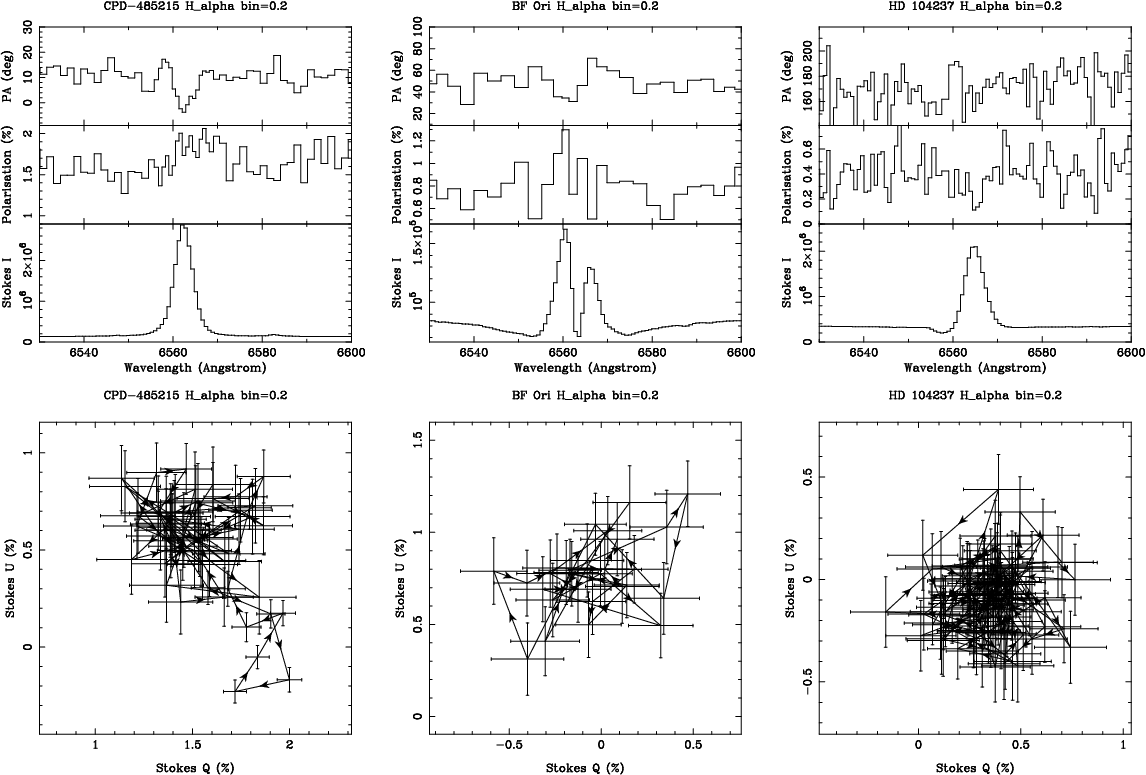}
       
   \caption{continued}

\end{figure*}

\subsubsection{Hydrogen recombination lines}

Spectropolarimetry was performed on all hydrogen lines including
H$\alpha$, H$\beta$ and the Paschen lines. The spectropolarimetric
data around H$\alpha$ are shown in a so-called ``triplot'' in the
upper half of Fig. \ref{halphafig}. In this triplot, the Stokes I
(normal intensity) is shown in the lower panel, the polarisation
percentage in the middle, while the position angle (PA) is shown in
the upper panel.  The results are also represented in a Stokes {(\it
  Q, U)} diagram (bottom) in Fig. \ref{halphafig} using the same
wavelength range of the triplot spectra, but sometimes with a
different binning. Ten out of the twelve objects display a clear line
effect across H$\alpha$ while the remaining two, BF Ori and HD 104237,
show a potential line effect. The observed line effects across
H$\alpha$ are of order of $\sim$0.2 \% to $\sim$7 \% in the case of R
Mon. The change in polarisation occurs across the emission lines, the
absorption components or the entire line profiles.

The spectropolarimetric properties of H$\alpha$ for each target are
listed in Table \ref{halpha}. Columns 3, 4 \& 5 list the spectroscopic
characterisation of Stokes I, the intensity spectrum. Generally, the
lines are stronger in HBe objects than in HAe objects. The final three
columns represent the line polarimetric properties, i.e. the
depolarisation, line polarisation and McLean effects as introduced in
Sec.~1.

The intrinsic polarisation angle is measured from the slope of the
loop of the line in the {(\it Q, U)} diagram following
\citet{2011A&A...532A..28W}. From the analysis of the {(\it Q, U)}
diagram, we can determine the direction of intrinsic polarisation on
the sky. We have measured the intrinsic polarisation from the slope of
the loop across the line in the {(\it Q, U)} diagram of
Fig. \ref{halphafig}. This gives two different results depending on
the type of the line effect. In the case of line depolarisation, the
intrinsic angle is measured from the line to continuum while for line
polarisation it is measured from continuum to the line resulting in a
$90^{\circ}$ difference from the former. This $90^{\circ}$ difference
in polarisation is equivalent to $180^{\circ}$ on the sky, for that
reason the direction of the vector in the {(\it Q, U)} diagram is
crucial, and this depends on the type of polarisation we observe.  For
the objects displaying the McLean effect, the angle is measured from
the continuum to the line.

\begin{figure}
        \centering
        \includegraphics[width=1\columnwidth]{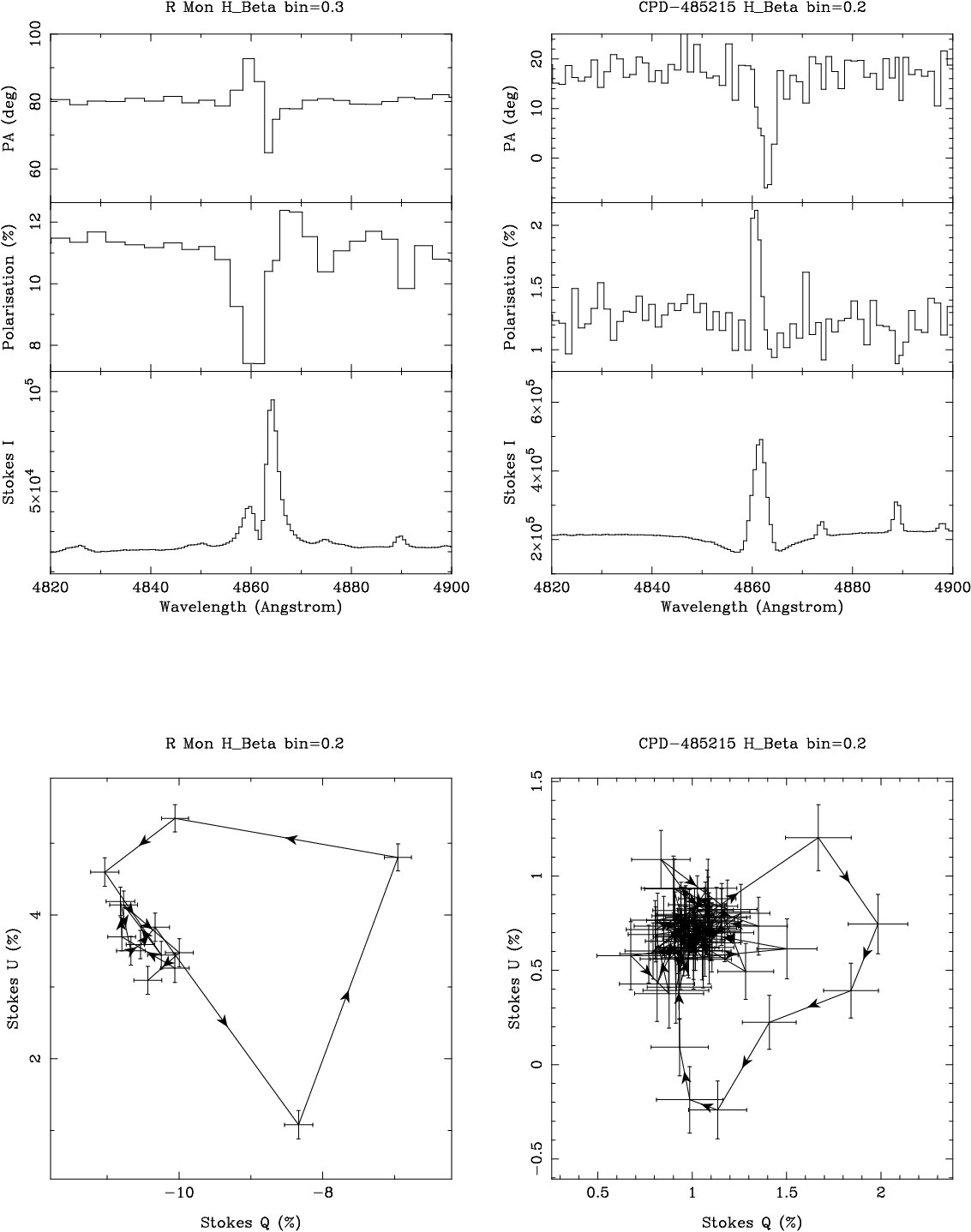}
       
   \caption{H$\beta$ spectropolarimetry of R Mon and CPD-485215. See the caption of Fig. \ref{halphafig} for more details. }
    \label{beta}
\end{figure}

The data for the next
strongest hydrogen recombination line, H$\beta$, is available for five
objects. Two of these, R Mon and CPD-485215, show a clear change in
polarisation across this line and the results are presented in
Fig. \ref{beta}. The Paschen lines are much weaker than the Balmer
series and only R Mon shows a line effect across these lines.

\begin{figure}
        \centering
        \includegraphics[width=1\columnwidth]{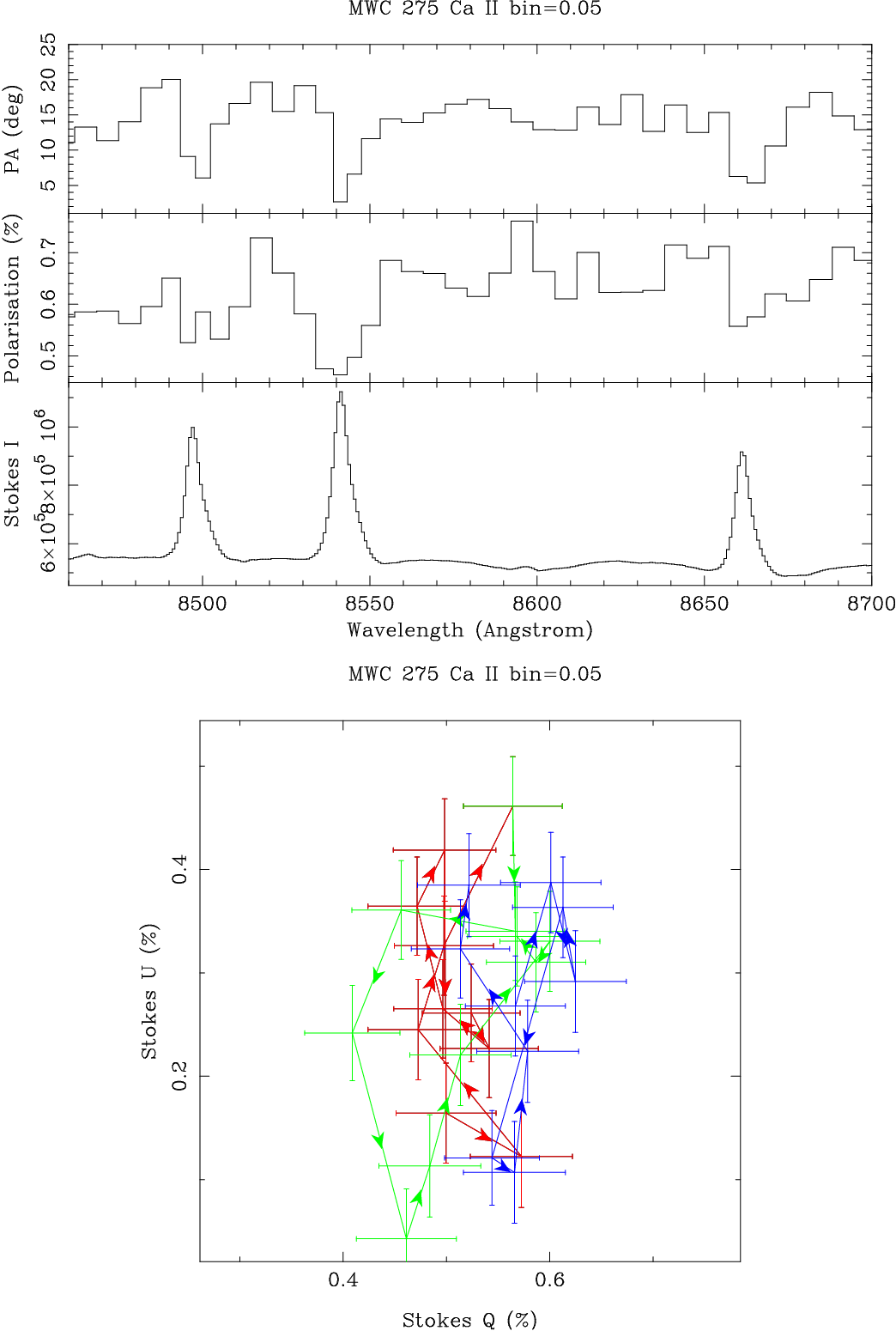}

   \caption{{Ca \sc ii} triplet polarisation data of MWC 275. In
  the {(\it Q, U)} diagram, the lines at 8498, 8542 and 8662~$\rm \AA$ are plotted in red, green and blue solid lines. See the caption of
     Fig. \ref{halphafig} for more details.}
    \label{mwc}
\end{figure}


\begin{table*}
 \centering
 \begin{minipage}{140mm}
\caption{The H$\alpha$ line results: columns (3), (4) and (5)
  list the Stokes (I) characteristics; columns (6), (7) and (8) list line
  spectropolarimetry characteristics of each target. Column (8)
  provides an estimation of the intrinsic polarisation PA derived from line excursions in
  the {(\it Q, U)} diagram (see text for details).  Errors in the EW
  measurements are typically 5\%, the uncertainty in intrinsic
  polarisation angle is approximately $10^{\circ}$.}
\label{halpha}
\begin{tabular}{l l l r r l l r}
\hline

  Object &  Spec. type    &line profile &  EW~($\rm\AA$)  &  Line/cont.  &  Line effect  &  Classification & $\theta_{intr}^{\circ}$   \\
\hline 
PDS 27 &    B3 &P Cygni&     -120.8   &     18.0   &     Yes       & McLean   &   77          \\
PDS 37  &     B3   &P Cygni&     -122.6   &       19.4 &     Yes        &  McLean&   56           \\
PDS 133&   B6 &P Cygni&   -94.5     &   20.7      &     Yes       & McLean  &   126             \\
HD 98922&   A0 &  P Cygni &  -17.9   &    6.0     &      Yes       & McLean$\backslash$Depol. & 22               \\
R Mon(2011)&     B8 &Double-peaked&  -98.6    &   12.6      &      Yes      & Depol. & 63             \\
R Mon(2012)&&Double-peaked&   -105.5    &  12.9      &     Yes        & Depol. &  57            \\
GU CMa &   B1 &Pure emission&   -11.1    &  2.6      &     Yes        & Depol. &  127         \\
CPD-485215&   B6  &Pure emission&  -100.8      &  19.6    &    Yes        & Depol. & 62          \\
MWC 275&   A1  &P Cygni&  -14.9     &  3.8     &     Yes        & $\sim$ Pol.  &  112         \\
HD 85567(2011) &   B7 &P Cygni& -43.5      &  9.9      &     Yes        & $\sim$ Depol.& 31         \\
HD 85567(2012)&&P Cygni&  -42.1     &  10.2     &     Yes        & $\sim$ Depol. &    22         \\
V380 Ori(2011)&   A0&Pure emission & -82.7      &  15.5    &      Possible   & &        \\
V380 Ori(2012)&& Pure emission&  -71.1     &  13.6    &     Yes   &    $\sim$ pol. &  10    \\
HD 104237   & A7  &P Cygni&  -27.2      &   6.5   &      Possible     &    &      \\
BF Ori  &   A2  &Double-peaked& 1.0       &   1.9   &      Possible     &  &     \\
\hline
\end{tabular}
\end{minipage}
\end{table*}

\subsubsection{{Ca \sc ii} lines}

The {Ca \sc ii} triplet at 8498, 8542 and 8662~$\rm \AA$ originates at
larger distances from the central star than H$\alpha$ because of its
lower ionisation energy. Only two objects out of twelve, show a change
in polarisation across these lines. MWC 275, as shown in
Fig. \ref{mwc}, displays a clear change across these lines. The change
is as broad as the emission lines and is clear in both polarisation
and polarisation angle. R Mon also shows a change in polarisation
across these lines (see Fig. \ref{all}) in appendix A).  PDS 37 shows
a possible line effect in only the first two datasets while the second
two datasets do not show such a line effect
\citep{ababakr15}. Therefore, the line effect in PDS 37 is not a
conclusive result based on the current data. The {Ca \sc ii} doublet
at 8912 and 8927~$\rm \AA$ is also seen in the spectra of most
objects. R Mon is the only object that shows a line effect across the {Ca
  \sc ii} doublet.

\subsubsection{{He \sc i} lines}

 The {He \sc i} lines trace the very inner hot circumstellar
 environment as they have a very high excitation energy. Three {He \sc i} lines at 5876, 6678 and 7065 ~$\rm \AA$ are seen
in the spectra of all objects. These lines
 do not show any signs of a line effect.  In most objects the lines are in
 absorption however, while those in emission are very weak, possibly
 too weak to show the line effect. 

\subsubsection{Forbidden lines}

\begin{figure}
       \centering
       \includegraphics[width=1\columnwidth]{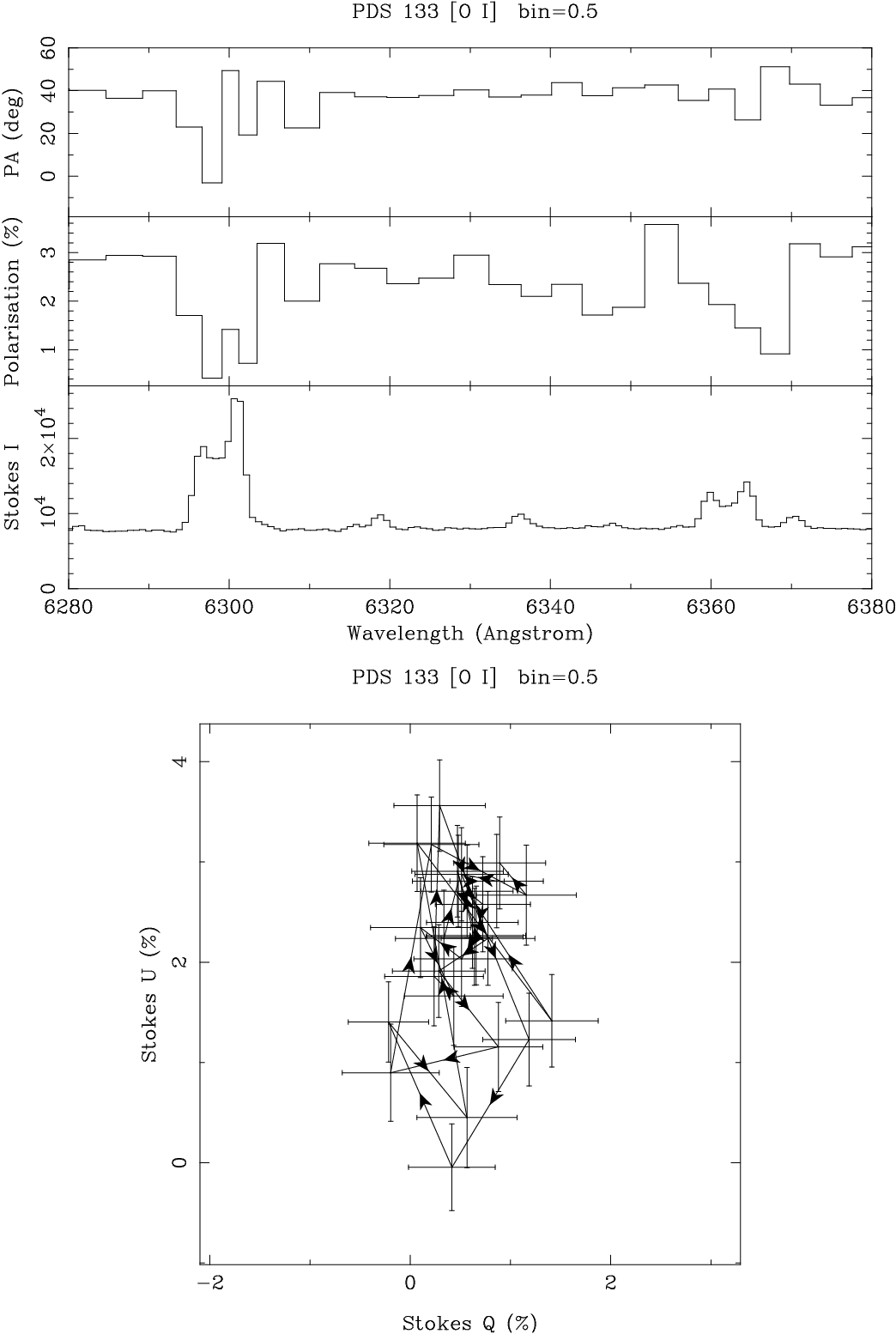}
      \caption{ {[O \sc i]} at 6300 and 6364~$\rm\AA$ polarisation
        data of PDS 133. See the caption of Fig. \ref{halphafig} for
        more details.}
      \label{133}
\end{figure}

Most objects have forbidden emission lines, however line effects
across them are limited to a few objects.  All the {[Fe \sc ii]} lines
show a depolarisation in R Mon and CPD-485215. Both {[S \sc ii]}
display a clear depolarisation in the spectra of R Mon. {[O \sc i]}
lines also display a clear depolarisation in R Mon, CPD-485215 and PDS
133. However, the lines are broad and double peaked in PDS 133. The
polarisation spectra of PDS 133 across {[O \sc i]} at 6300~$\rm \AA$
is presented in Fig. \ref{133} and the remainder of the polarisation
spectra of all forbidden lines are presented in Fig. \ref{all}.

\subsubsection{O I and Fe II lines}

Two {O \sc i} lines at 7773 and 8446 $\rm \AA$ are observable. R Mon and CPD-485215 show a clear change in polarisation across these lines. Several {Fe \sc ii} lines are observable in the spectra of all objects. However, only R Mon shows the line effect across these lines. The polarisation spectra can be seen in Fig. \ref{all}.

\subsubsection{Summary on the occurrence of the line effect}

We presented spectropolarimetry from 4560~$\rm\AA$ to 9480~${\rm \AA}$ of a
sample of HAeBe stars that were previously known to have a line
effect across the H$\alpha$ line. We find that this phenomenon is very
rare across lines other than H$\alpha$. Strong emission lines such as
H$\beta$ and the Ca{\sc ii} near-infrared triplet are only seen with a
line effect in just a few sources. We should perhaps note here that
the detection does not necessarily depend on the line strength. For
example, lines such as H$\beta$ and even H$\gamma$ in the Be star
$\zeta$ Tau have line effects of order 1\%, while the lines themselves
hardly reach above the continuum (see the review by \citealt{oudmaijer07}).
The fact that strong lines do not exhibit a line effect is thus a very
strong observational finding. Finally, R Mon appears to be an
exceptional object in the sense that it is the only star in our sample
that has a polarisation effect in most of its emission lines. We will
return to this object later in the following discussion.

\section{Analysis}

The spectropolarimetric results of the sample show a continuum
polarisation at the level between $\sim$0.3 \% to $\sim$12 \%. Unlike
the polarised standard stars, the majority of objects do not follow
the empirical curve of the Serkowski law. A line
effect is present across the H$\alpha$ line in all objects.



We observe three different line effects across the H$\alpha$ line in
the sample (see Table \ref{halpha}): depolarisation, intrinsic
polarisation; and a polarisation change across the absorption, which
is commonly referred to as the McLean effect. In depolarisation, the
polarisation across the emission line decreases compared to the
continuum, and it is more common in high mass stars
\citep{1974MNRAS.167P..27C, 1976ApJ...206..182P}.  Intrinsic line
polarisation can be due to an anisotropic line emitting region
scattering off circumstellar material, and it was found in T Tauri and HAe stars \citep{Vink2002, vink03, vink05a}. The McLean
effect is where the absorption component of the emission line often
has a different polarisation than the continuum, but the emission has
not \citep{McLean1979}. There is a distinct difference in polarimetric
properties between the HBe stars, showing de-polarisation and
the HAe stars showing enhanced polarisation across the line.

The intrinsic line polarisation is thought to be caused by the effects
of magnetospheric accretion. In this case, disk material is funnelled
through accretion columns on to the stars. The material shocks the
photosphere where it crashes into at high speeds.  The line emission coming
from these hot spots will be polarised as they scatter off the disk
material, which becomes detectable as a line effect \citep*{vink05b}.

\begin{figure*}
        \centering
        \includegraphics[width=16cm]{absorption.ps}
        
   \caption{H$\alpha$ and H$\beta$ spectropolarimetry of PDS 27, PDS
     37 and PDS 133. For PDS 133 only H$\alpha$ data is available. See
     the caption of Fig. \ref{halphafig} for more details.}
    \label{config}
\end{figure*}

\begin{figure*}
        \centering
        \includegraphics[width=16cm]{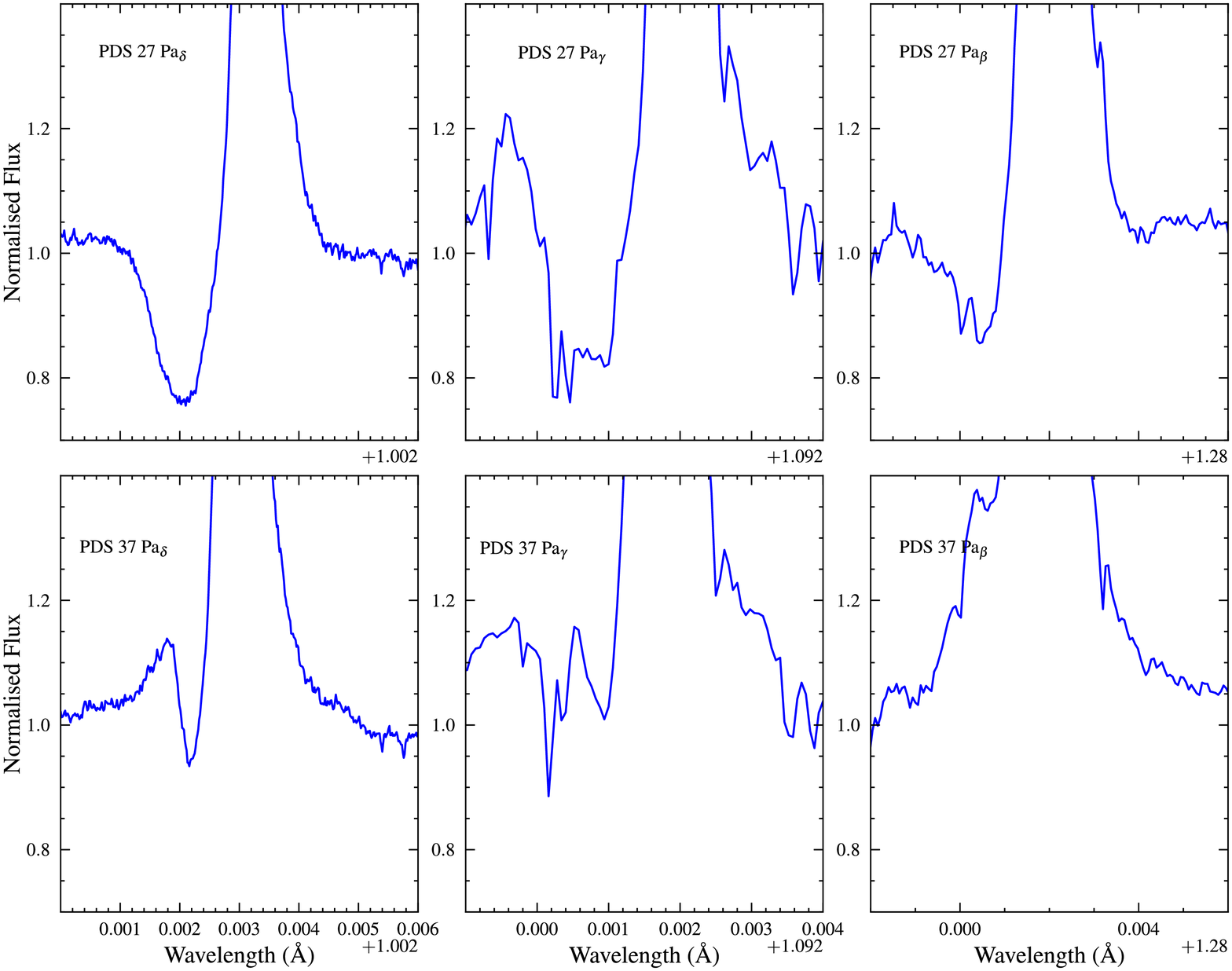}
        
   \caption{The comparison of P Cygni profile,Pa$\delta$, Pa$\gamma$
     and Pa$\beta$ of PDS 27 and PDS 37. The spectroscopic data are
     taken from the HAeBe X-shooter project \citep{Oudmaijer11,
       Fairlamb2015}}
    \label{config1}
\end{figure*}

\subsection{McLean effect}

A number of objects display a polarisation change across the
P Cygni absorption of the hydrogen lines and do not necessarily fit into the
categories above.

The concept of the McLean effect is based on the fact that the 
wind material in our line of sight blocks the unscattered light and
removes it from the beam. This leads to observation of a strong
emission line with a blueshifted absorption component. However, the
flux in the absorption does not reach zero because the re-emission
process is isotropic; some of the photons will be scattered into our
line of sight. If the distribution of the scattering material is
aspherical, the observed light across the absorption will be more
polarised than the continuum light.
PDS 27, PDS 37 and PDS 133 show a clear McLean effect across the
absorptive component of H$\alpha$. The line effect is of order 1-2\%
in these objects. Naively, we might expect the same line effect across
the H$\beta$ but this is not observed.
Fig. \ref{config} shows the H$\alpha$ and H$\beta$ P Cygni profile of
PDS 27, PDS 37 and PDS 133. The P Cygni absorption seems to be
saturated with broad and flat core features for PDS 27 and the
absorption is around 0.4 and 0.15 of the continuum level for H$\alpha$
and H$\beta$ respectively. In PDS 37, the P Cygni absorption is at
0.75 and 0.3 of the continuum level for H$\alpha$ and H$\beta$
respectively. The absorption in H$\beta$ is thus deeper than that of
H$\alpha$ in both objects. This is not expected as typically the
higher lines display weaker absorption. This unusual behaviour is also
seen for the Paschen lines as presented in Fig.  \ref{config1}. The
figure shows that the absorption component of Pa$\beta$ is weaker than
Pa$\gamma$ and Pa$\delta$.  Unlike PDS 27 and PDS 37 the absorption
part of the line of PDS 133 is above the continuum.

By analysing the observed P Cygni spectral lines and the observed
polarisation line effects, we can understand the McLean effect in
H$\alpha$ and its absence in H$\beta$ relatively straightforwardly. The
strong H$\alpha$ line photons are scattered into the line of sight
filling in the absorption component. If the scattering region is
a-symmetric, net polarisation across the absorption part of the line
will be observable.

The H$\beta$ line, which is weaker than H$\alpha$ has fewer photons, and
will therefore also have fewer scattered photons filling the
absorption component (see Fig. \ref{config}). The net effect is that
both the H$\beta$ absorption remains deeper than for H$\alpha$, while
the scattered, polarised component is less obvious.  Therefore, the
lack of the scattered photons in H$\beta$ could explain the absence of
the line effect across this line. This argument is supported by the
observed line effects across the absorptive component of H$\alpha$ in
PDS 27, PDS 37 and PDS 133, as the line effect is stronger for a
weaker absorption component. \cite{ababakr15} suggested a similar
scenario in an in-depth study of PDS 27 and PDS 37, but here we have
an extra object that follows the same pattern.

\begin{figure}
       \centering
       \includegraphics[width=1\columnwidth]{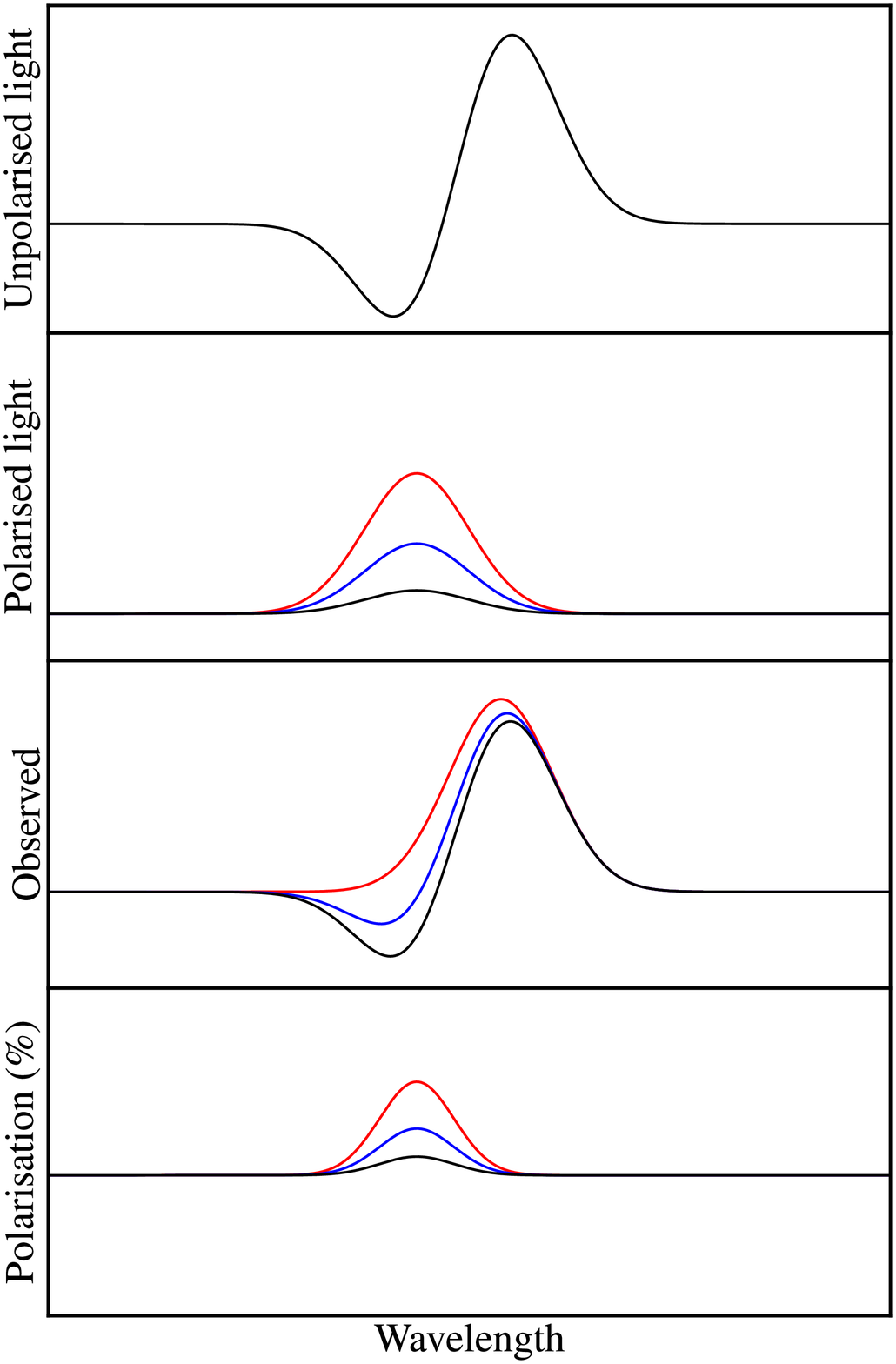}
       \caption{ Schematic explaining the McLean effect with the data
         present in three sets: the unpolarised light is shown in the
         upper panel, polarised light into the line of sight in the
         upper middle, the observed H$\alpha$ line profile  in the
         lower middle and the observed polarisation (\%) is shown in
         the lower panel. The absorption is completely filled when
         there is a significant amount of polarisation in our line of
         sight, this results in a detection of a notable line effect
         (red solid line). In contrast, when the polarised light into
         our line of sight is not sufficient to fill in the absorption
         component significantly, this results in a weak line effect
         which is challenging to observe (black solid line). The solid
         blue line represents an average between the two extreme
         cases.}
       \label{model}
\end{figure}

For illustration, we sketch the situation in Fig. \ref{model}. In the
schematic we can see that different levels of polarisation in the line
of sight result in different absorption line profiles. In the case of
large polarisation, we do not see any absorption in the observed line
profile as the polarised light fills in the absorption. Interestingly,
the McLean effect can thus also be observed in the absence of 
absorption. In contrast, very deep absorption would be observed when
there is little or no polarisation.

\subsection{The case of R Mon}

A small number of objects display line effects across more lines than
H$\alpha$ alone. We discuss these here, focussing on the most extreme
object, R Mon.  Its continuum polarisation is large, with values of
order 12\% and 8\%, in the {\it R} band in 2012 and 2011 respectively.
Unlike the other stars, R Mon displays a line effect across virtually
all observed emission lines. The difference between line and continuum
polarisation is 7\% across the H$\alpha$ and H$\beta$ lines (see
Figs. \ref{halphafig} and \ref{beta}), which is large and
rare. Typically, the strongest line effect due to electron scattering
is of order 2\% \citep{oudmaijer07}. In addition, the direction of the intrinsic polarisation angle from the line effects
  in the {(\it Q, U)} diagram is not constant over different
  bands.
Therefore, it would appear that R Mon is atypical as it differs on
many counts from most objects that have been studied
spectropolarimetrically.

Relevant to this discussion is that R Mon is an extended object. Its
reflection nebula is due to radiation from the inner parts, such as
the star and wind, being scattered off circumstellar dust particles,
resulting in strong polarisation. Spatially resolved polarisation
images reveal polarisation in excess of 50 \% \citep{jolin10}. The polarised flux is
largest closest to the star and decreases with distance, while the
polarisation angles follow a centro-symmetric pattern. For a
circularly symmetric, highly polarised object on the sky, the same
would apply, but if it were unresolved, the net observed polarisation
would be zero. However, because R Mon is asymmetric, a net
polarisation will be observed, and this is even stronger in our data
as the scattering circumstellar material is resolved and larger than
the slit.

These features have important consequences for the observed
polarimetry. Below we will argue that this is the main reason that the
(spectro-)polarimetric properties of R Mon are different from the
other stars, and that the various emission lines originate from
different line forming regions.  For example, given that the
polarisation drops off rapidly with distance from the star, a slight
change in seeing or slight change in the positioning of the slit can
result in largely different polarised light entering the slit. In
addition, larger positioning errors can result in differently
observed position angles, as the PA depends on position.

Inspection of the raw spectropolarimetric data shows that the width of
the ordinary (O) and extraordinary (E) beam in the spatial direction
in the 2011 data is about twice that of the 2012 data (and wider than
the data for the other stars). In addition, the individual beams do
not follow the typical Gaussian shape of the seeing profile, but
appear asymmetric. This confirms that R Mon is not a point source but
an irregular, extended object.

Let us discuss the polarisation properties of R Mon in this context,
starting with its observed broadband peculiarities. The polarisation
seemingly increased by 4\% over a timespan of 4 months, while the
polarisation angle stayed the same.
The slit direction and position were the same for both observations
but as mentioned above, the spatial profile of the 2011 data, which
has a smaller level of polarisation, is much wider than that in
2012. As the polarisation closest to the object is highest, it follows
that the net polarisation in 2012 is higher. A similar argument can be
put forward to explain the discrepant position angles for each of the
wavebands the object was observed. In Fig. \ref{allcon}, the slope of
the PA is the same for the $V$, $R$ and $I$ bands, but they are offset
by 10$^{\rm o}$ (see also Table \ref{con}). According to the data in
\citet{jolin10}, the polarisation is centro-symmetric, and the
polarisation angle changes with at least 20$^{\circ}$ every 1
arcsec. Hence, a slight displacement can result in a noticeable change
in PA.  We argue, therefore, that the fact that R Mon is extended and
asymmetric explains the high polarisation, large continuum
variability, and offset in PA for the various wavelength settings.

Let us now turn to the spectropolarimetric properties, of which the
fact that most lines seem to exhibit a line effect, and that the line
effect is large, makes R Mon stand out.

\begin{figure*}
        \centering
        \includegraphics[width=16cm]{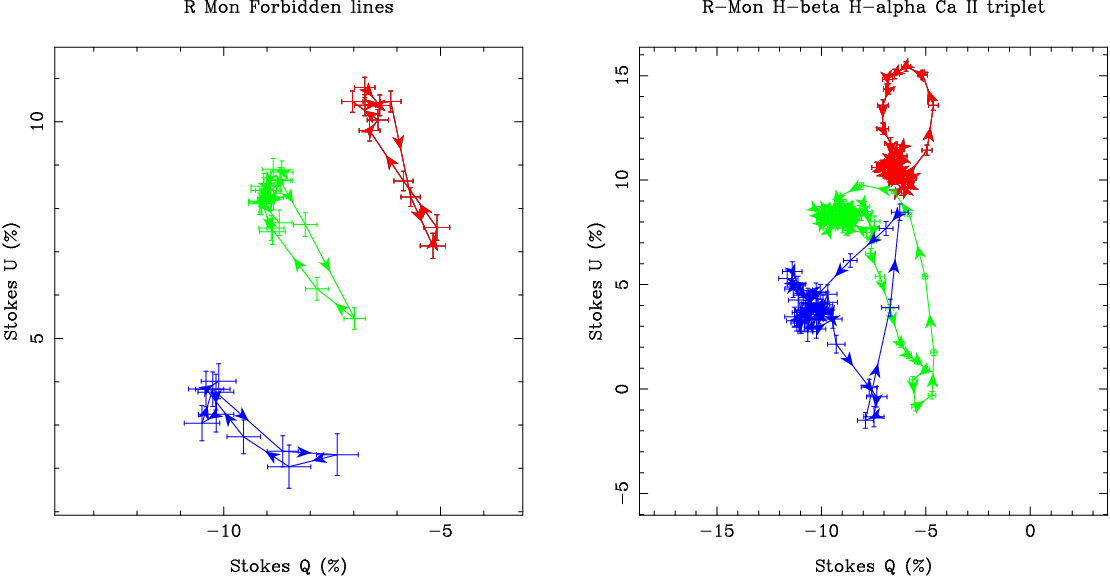}
        
   \caption{{(\it Q, U)} diagram of several spectral lines and forbidden lines of R Mon. The lines are taken from {\it B} band (blue solid line), {\it R} band (green solid line) and {\it I} band (red solid line). H$\beta$,  H$\alpha$ and {Ca \sc ii} triplet at 8662~$\rm \AA$ are presented in blue, green and red lines respectively (right); three forbidden lines: {[Fe \sc ii]} at 5158~$\rm \AA$, {[O \sc i]} at 6364~$\rm \AA$ and {[Fe \sc ii]} at 8616 $\rm \AA$ are shown in blue, green and red lines respectively (left).}
    \label{qurmon}
\end{figure*}

The left hand panel of Fig. ~\ref{qurmon} shows the {(\it Q, U)}
diagram of three forbidden lines of R Mon, with each line taken from
a different wavelength setting. The right hand panel shows lines
from different wavelength settings as well; H$\beta$, H$\alpha$ and a
{Ca \sc ii} triplet line, at 8662 $\rm \AA$. The line excursions are large,
ranging from $\sim$5\% to $\sim$10\% in H$\alpha$.
The direction of the intrinsic polarisation angle is roughly the same
for all forbidden lines, with an average value of
64~$\pm$~$5^{\circ}$. However, very different intrinsic angles of
90~$\pm$~$15^{\circ}$, 57~$\pm$~$10^{\circ}$ and
128~$\pm$~$10^{\circ}$ are found for H$\beta$, H$\alpha$ and the Ca
{\sc ii} triplet line respectively. The H$\alpha$ and Ca
{\sc ii} triplet data
were taken at the same date while H$\beta$ was taken a few days
later. The polarisation angle in H$\alpha$ is the only one consistent
with the forbidden lines.

The question that now arises is how can the line effects be strong,
and why would the intrinsic angles be different for the various lines?
By simply stating that the lines are formed in a larger region than
the star and would be less prone to scattering, such as the
polarisation dilution discussed in \citet{trammell94}, does not
hold in the current situation. The reflecting dust is located far away
from the star and is in all likelihood much more extended than the
ionized wind. In this case, the emission lines will be equally likely
to be scattered and thus as polarised as the continuum. The remaining
possibility is that the line emitting regions themselves are not
isotropic - either asymmetric or clumpy, leading to a polarisation
line emission different from the continuum. The different polarisation
angles for the various lines may be the result of different geometries
and volumes of the line forming regions. The optically thick H$\alpha$
emission and the forbidden lines originate from large, possibly
similar volumes, which could follow similar scattering paths, whereas
other lines arise in regions closer to the star and have potentially
different geometries, leading to different polarisation angles.

To our knowledge this is the first time an object has been reported to
have so many lines with a polarisation effect across them.  It will not
be trivial to disentangle the various contributions and
multi-parameter modelling is needed to retrieve the geometry of the
system. This is beyond the scope of the current paper.

\section{Final Remarks and Conclusions}
 
We have performed the first linear spectropolarimetry over a large
wavelength range of a sample of HAeBe stars. The spectra cover
the range from 4560~$\rm\AA$ to $9480~{\rm \AA}$, including many
emission lines such as: hydrogen recombination lines, {Ca \sc ii}, {Fe
  \sc ii}, {O\sc i} and {He \sc i} and several forbidden lines
including {[O \sc i]}, {[Fe \sc ii]} and {[S \sc ii]}.  All the
objects have a measurable continuum polarisation, ranging from
$\sim$0.3 \% in the very low polarised star HD 85567 to $\sim$12 \% in
the highly polarised star R Mon.  Our values are broadly consistent
with these values but there are some variations from the literature
values. The continuum polarisation of some of our objects do not
follow the general trend of the Serkowski law. This is evidence of the
presence of intrinsic polarisation in the observed continuum
polarisation as the interstellar polarisation does not vary
significantly over time.

    
All the objects show a sign of line effect across H$\alpha$. All HBe
objects display either a classical depolarisation or a McLean effect
signature. The HAe objects show a mixture of McLean, depolarisation,
intrinsic line polarisation and complex line effect. This suggests
that the circumstellar environments around HAeBe stars have a
flattened structure. Four objects in our sample show a McLean effect,
where the change in polarisation occurs across the absorptive component of
the emission line of H$\alpha$.
This suggests that the distribution of the ionised material is
not circular on the sky. From these four objects we only have H$\beta$
data for PDS 27 and PDS 37. We would expect the same line effect
across the absorption component of H$\beta$.
However, the absorption in H$\beta$ is approximately three times
stronger than that of H$\alpha$.
We can explain this apparent inconsistency by the fact that strong
H$\alpha$ line emission scatters into our line of sight filling in the
absorption component. The H$\beta$ line is weaker than H$\alpha$ and
thus fewer scattered photons fill in the absorption component,
yielding a much lower polarisation effect, consistent with the
observed non-detection.

It would appear, therefore, that we can explain the differing
  polarisation behaviour across the absorption parts of the H$\alpha$
  and H$\beta$ lines very well in the context of the McLean effect. An
  alternative scenario to explain polarisation across the absorption
  parts of hydrogen lines was put forward by \citet{kuhn07}. Essentially this invokes selective optical pumping of
  the lines, predicting that both H$\beta$ and H$\alpha$ will have polarisation
  effects of roughly similar magnitude (\citealt{kuhn11}, their Figs. 4 and 5). This is not what we
  observe in our data. What is more, this scenario will also struggle
  to explain why the H$\beta$ absorption is so much stronger than that
  of H$\alpha$. Within the current set of observations,  the McLean
  effect seems to be a more viable explanation.

Four HBe objects show a depolarisation line effect, where the change
in polarisation is as broad as the emission line. This suggests that
H$\alpha$ emerge in an extended circumstellar disk. Two HAe objects
objects show an intrinsic line polarisation, where a significant
portion of the photons emerge from accretion hot spots. BF Ori and HD
104237 show a possible line effect across H$\alpha$.

We have observed depolarisation line effects across {[O \sc i]} lines
at 6300 and 6364~$\rm\AA$ in PDS 133, R Mon and CPD-485215. The line
profile in PDS 133 is broad and double peaked unlike the other two,
which are narrow and single peaked. The line profiles may suggest that the
lines originate from a rotating disk in PDS 133
while in R Mon and CPD-485215 many forbidden lines display a
depolarisation line effect. It is to be expected to see a depolarisation
line effect across {[O \sc i]} in PDS 133 as they originate in the
disk further out from the inner ionised disk. The {Ca \sc ii} triplet also
shows a depolarisation in MWC 275. These lines have low ionisation
potential energy as they are formed in a predominately neutral zone. In
general they originate in the circumstellar environment but outside
the hydrogen ionisation boundary. Therefore, we might  expect a
depolarisation line effect across these lines. However, for the
objects that display a depolarisation line effect, the {Ca \sc ii} triplet
line effects are either absent or very weak.

Finally, we have presented the medium resolution spectropolarimetry on the
entire optical range from 4560~$\rm\AA$ to $9480~{\rm \AA}$ of a
sample of 12 HAeBe stars. The data allow us to draw the following
conclusions:

\begin{itemize}

\item We sample linear line spectropolarimetry in the optical
  wavelength range, which is much larger than any previous work at
  similar spectral resolution.

\item Changes in the polarisation across the H$\alpha$ emission line
  are detected in all objects, as an indication of a flattened
  structure of the circumstellar environment. The line effects vary
  from depolarisation, line polarisation to the McLean effect.

\item Depolarisation and the McLean effect are observed in Be type
  stars predominantly in early B type stars while line polarisation is
  observed in Ae type objects.

 \item The McLean effect is observed only across the absorptive
   component of H$\alpha$ and the line effect is stronger for a weaker
   absorption component, while H$\beta$ does not display the
   effect. We propose a scenario to explain this property.  It is
   based on the fact that the photons from the strong H$\alpha$ line
   are scattered into the line of sight. As a consequence, the photons
   in the absorption are more polarised than the emission. A side
     result is that the selective absorption due to optical pumping as
     proposed by \citet{kuhn07} is unlikely to be responsible for the
     polarisation behaviour in these objects.

 \item We detect a broad depolarisation line effect across {Ca~\sc ii}
   triplet and {[O \sc i]} in two objects. These lines are emerging
   further away from the star and H$\alpha$ region in the
   circumstellar environment. The depolarisation simply implies that
   the circumstellar environment has an asymmetrical structure in this
   region.

 \item The few spectra with calcium triplet lines that show
   an effect show a similar polarisation profile for all members of
   the triplet. We confirm a similar observation by \citet{kuhn11} for an evolved, RV Tau star. As these authors
   explained, resonant line scattering can then not be the cause for
   the observed line polarisation, as the {Ca~\sc ii} 8662~$\rm \AA$ comes from
   a different upper level than the other members of the triplet.
   Similar polarisation due to line scattering would then not be
   expected.

\item Finally, apart from H$\alpha$, few lines in few objects show a line
   effect in the polarisation. We present the case of R Mon, which
   displays exceptionally strong line effects in most of its emission
   lines. In addition, not all lines show the same type of line
   effect. We explain this by the fact that the object itself is
   resolved and propose that different line forming regions and
   geometries are responsible for polarisation properties of the
   higher hydrogen recombination lines and forbidden lines and
   H$\alpha$ on the other hand.
   
 \end{itemize}

\section*{Acknowledgements}

We are very grateful to the referee, Richard Ignace, for his
constructive and very insightful remarks that helped us improve the
paper.  This research has made use of the SIMBAD database, operated at
CDS, Strasbourg, France.

\bibliographystyle{mn2e}

\bibliography{ref.bib}

\begin{thebibliography}{}

\bibitem[\protect\citeauthoryear{{Ababakr}, {Fairlamb}, {Oudmaijer} \& {van den
  Ancker}}{{Ababakr} et~al.}{2015}]{ababakr15}
{Ababakr} K.~M.,  {Fairlamb} J.~R.,  {Oudmaijer} R.~D.,    {van den Ancker}
  M.~E.,  2015, \mnras, 452, 2566

\bibitem[\protect\citeauthoryear{{Acke}, {van den Ancker} \&
  {Dullemond}}{{Acke} et~al.}{2005}]{acke05}
{Acke} B.,  {van den Ancker} M.~E.,    {Dullemond} C.~P.,  2005, \aap, 436, 209

\bibitem[\protect\citeauthoryear{{Beltr{\'a}n} \& {de Wit}}{{Beltr{\'a}n} \&
  {de Wit}}{2016}]{beltran16}
{Beltr{\'a}n} M.~T.,  {de Wit} W.~J.,  2016, \aapr, 24, 6

\bibitem[\protect\citeauthoryear{{Clarke} \& {McLean}}{{Clarke} \&
  {McLean}}{1974}]{1974MNRAS.167P..27C}
{Clarke} D.,  {McLean} I.~S.,  1974, \mnras, 167, 27P

\bibitem[\protect\citeauthoryear{{Fairlamb}, {Oudmaijer}, {Mendigut{\'{\i}}a},
  {Ilee} \& {van den Ancker}}{{Fairlamb} et~al.}{2015}]{Fairlamb2015}
{Fairlamb} J.~R.,  {Oudmaijer} R.~D.,  {Mendigut{\'{\i}}a} I.,  {Ilee} J.~D.,
   {van den Ancker} M.~E.,  2015, \mnras, 453, 976

\bibitem[\protect\citeauthoryear{{Fossati}, {Bagnulo}, {Mason} \& {Landi
  Degl'Innocenti}}{{Fossati} et~al.}{2007}]{Fossati07}
{Fossati} L.,  {Bagnulo} S.,  {Mason} E.,    {Landi Degl'Innocenti} E.,  2007,
  in {Sterken} C.,  ed., The Future of Photometric, Spectrophotometric and
  Polarimetric Standardization Vol.~364 of Astronomical Society of the Pacific
  Conference Series, {Standard Stars for Linear Polarization Observed with
  FORS1}.
p.~503

\bibitem[\protect\citeauthoryear{{Grady}, {Fukagawa}, {Maruta}, {Ohta},
  {Wisniewski}, {Hashimoto}, {Okamoto}, {Momose} et~al.,}{{Grady}
  et~al.}{2015}]{grady15}
{Grady} C.,  {Fukagawa} M.,  {Maruta} Y.,  {Ohta} Y.,  {Wisniewski} J.,
  {Hashimoto} J.,  {Okamoto} Y.,  {Momose}   et~al., 2015, \apss, 355, 253

\bibitem[\protect\citeauthoryear{{Grinin}}{{Grinin}}{1994}]{1994ASPC...62...63%
G}
{Grinin} V.~P.,  1994, in {The} P.~S.,  {Perez} M.~R.,   {van den Heuvel}
  E.~P.~J.,  eds, The Nature and Evolutionary Status of Herbig Ae/Be Stars
  Vol.~62 of Astronomical Society of the Pacific Conference Series,
  {Polarimetric activity of Herbig Ae/Be stars}.
p.~63

\bibitem[\protect\citeauthoryear{{Harries}}{{Harries}}{1996}]{Harries96}
{Harries} T.~J.,  1996, Starlink User Note, 204

\bibitem[\protect\citeauthoryear{{Herbig}}{{Herbig}}{1960}]{1960ApJS....4..337%
H}
{Herbig} G.~H.,  1960, \apjs, 4, 337

\bibitem[\protect\citeauthoryear{{Hillenbrand}, {Strom}, {Vrba} \&
  {Keene}}{{Hillenbrand} et~al.}{1992}]{1992ApJ...397..613H}
{Hillenbrand} L.~A.,  {Strom} S.~E.,  {Vrba} F.~J.,    {Keene} J.,  1992, \apj,
  397, 613

\bibitem[\protect\citeauthoryear{{Ignace} \& {Brimeyer}}{{Ignace} \&
  {Brimeyer}}{2006}]{ignace06}
{Ignace} R.,  {Brimeyer} A.,  2006, \mnras, 371, 343

\bibitem[\protect\citeauthoryear{{Ilee}, {Fairlamb}, {Oudmaijer},
  {Mendigut{\'{\i}}a}, {van den Ancker}, {Kraus} \& {Wheelwright}}{{Ilee}
  et~al.}{2014}]{ilee14}
{Ilee} J.~D.,  {Fairlamb} J.,  {Oudmaijer} R.~D.,  {Mendigut{\'{\i}}a} I.,
  {van den Ancker} M.~E.,  {Kraus} S.,    {Wheelwright} H.~E.,  2014, \mnras,
  445, 3723

\bibitem[\protect\citeauthoryear{{Jain} \& {Bhatt}}{{Jain} \&
  {Bhatt}}{1995}]{Jain95}
{Jain} S.~K.,  {Bhatt} H.~C.,  1995, \aaps, 111, 399

\bibitem[\protect\citeauthoryear{{Jolin}, {Bastien}, {Denni}, {Lafreni{\`e}re},
  {Doyon} \& {Voyer}}{{Jolin} et~al.}{2010}]{jolin10}
{Jolin} M.-A.,  {Bastien} P.,  {Denni} F.,  {Lafreni{\`e}re} D.,  {Doyon} R.,
   {Voyer} P.,  2010, \apj, 721, 1748

\bibitem[\protect\citeauthoryear{{Kraus}}{{Kraus}}{2015}]{kraus15}
{Kraus} S.,  2015, \apss, 357, 97

\bibitem[\protect\citeauthoryear{{Kuhn}, {Berdyugina}, {Fluri}, {Harrington} \&
  {Stenflo}}{{Kuhn} et~al.}{2007}]{kuhn07}
{Kuhn} J.~R.,  {Berdyugina} S.~V.,  {Fluri} D.~M.,  {Harrington} D.~M.,
  {Stenflo} J.~O.,  2007, \apjl, 668, L63

\bibitem[\protect\citeauthoryear{{Kuhn}, {Geiss} \& {Harrington}}{{Kuhn}
  et~al.}{2011}]{kuhn11}
{Kuhn} J.~R.,  {Geiss} B.,    {Harrington} D.~M.,  2011, in {Kuhn} J.~R.,
  {Harrington} D.~M.,  {Lin} H.,  {Berdyugina} S.~V.,  {Trujillo-Bueno} J.,
  {Keil} S.~L.,   {Rimmele} T.,  eds, Solar Polarization 6 Vol.~437 of
  Astronomical Society of the Pacific Conference Series, {Using Absorptive
  Linear Polarization Spectroscopy to Understand Imbedded Stars}.
p.~245

\bibitem[\protect\citeauthoryear{{McLean}}{{McLean}}{1979}]{McLean1979}
{McLean} I.~S.,  1979, \mnras, 186, 265

\bibitem[\protect\citeauthoryear{{Mottram}, {Vink}, {Oudmaijer} \&
  {Patel}}{{Mottram} et~al.}{2007}]{Mottram07}
{Mottram} J.~C.,  {Vink} J.~S.,  {Oudmaijer} R.~D.,    {Patel} M.,  2007,
  \mnras, 377, 1363

\bibitem[\protect\citeauthoryear{{Muzerolle}, {Calvet} \&
  {Hartmann}}{{Muzerolle} et~al.}{1998}]{1998ApJ...492..743M}
{Muzerolle} J.,  {Calvet} N.,    {Hartmann} L.,  1998, \apj, 492, 743

\bibitem[\protect\citeauthoryear{{Oudmaijer}}{{Oudmaijer}}{2007}]{oudmaijer07}
{Oudmaijer} R.~D.,  2007, in Hartquist T. W., Falle S. A. E. G., Pittard J. M.,
  eds, Diffuse Matter FromStar Forming Regions to Active Galaxies.
  Springer-Verlag, Berlin, p. 83

\bibitem[\protect\citeauthoryear{{Oudmaijer} \& {Drew}}{{Oudmaijer} \&
  {Drew}}{1999}]{1999MNRAS.305..166O}
{Oudmaijer} R.~D.,  {Drew} J.~E.,  1999, \mnras, 305, 166

\bibitem[\protect\citeauthoryear{{Oudmaijer}, {Palacios}, {Eiroa}, {Davies},
  {de Winter} et~al.,}{{Oudmaijer} et~al.}{2001}]{2001A&A...379..564O}
{Oudmaijer} R.~D.,  {Palacios} J.,  {Eiroa} C.,  {Davies} J.~K.,  {de Winter}
  D.,    et~al., 2001, \aap, 379, 564

\bibitem[\protect\citeauthoryear{{Oudmaijer}, {Proga}, {Drew} \& {de
  Winter}}{{Oudmaijer} et~al.}{1998}]{Oudmaijer1998}
{Oudmaijer} R.~D.,  {Proga} D.,  {Drew} J.~E.,    {de Winter} D.,  1998,
  \mnras, 300, 170

\bibitem[\protect\citeauthoryear{{Oudmaijer}, {van den Ancker}, {Baines},
  {Caselli}, {Drew}, {Hoare}, {Lumsden}, {Montesinos}, {Sim}, {Vink},
  {Wheelwright} \& {de Wit}}{{Oudmaijer} et~al.}{2011}]{Oudmaijer11}
{Oudmaijer} R.~D.,  {van den Ancker} M.~E.,  {Baines} D.,  {Caselli} P.,
  {Drew} J.~E.,  {Hoare} M.~G.,  {Lumsden} S.~L.,  {Montesinos} B.,  {Sim} S.,
  {Vink} J.~S.,  {Wheelwright} H.~E.,    {de Wit} W.~J.,  2011, Astronomische
  Nachrichten, 332, 238

\bibitem[\protect\citeauthoryear{{Poeckert} \& {Marlborough}}{{Poeckert} \&
  {Marlborough}}{1976}]{1976ApJ...206..182P}
{Poeckert} R.,  {Marlborough} J.~M.,  1976, \apj, 206, 182

\bibitem[\protect\citeauthoryear{{Quirrenbach}, {Buscher}, {Mozurkewich},
  {Hummel} \& {Armstrong}}{{Quirrenbach} et~al.}{1994}]{quirrenbach94}
{Quirrenbach} A.,  {Buscher} D.~F.,  {Mozurkewich} D.,  {Hummel} C.~A.,
  {Armstrong} J.~T.,  1994, \aap, 283, L13

\bibitem[\protect\citeauthoryear{{Rodrigues}, {Sartori}, {Gregorio-Hetem} \&
  {Magalh{\~a}es}}{{Rodrigues} et~al.}{2009}]{2009ApJ...698.2031R}
{Rodrigues} C.~V.,  {Sartori} M.~J.,  {Gregorio-Hetem} J.,    {Magalh{\~a}es}
  A.~M.,  2009, \apj, 698, 2031

\bibitem[\protect\citeauthoryear{{Serkowski}, {Mathewson} \&
  {Ford}}{{Serkowski} et~al.}{1975}]{1975ApJ...196..261S}
{Serkowski} K.,  {Mathewson} D.~S.,    {Ford} V.~L.,  1975, \apj, 196, 261

\bibitem[\protect\citeauthoryear{{The}, {de Winter} \& {Perez}}{{The}
  et~al.}{1994}]{1994A&AS..104..315T}
{The} P.~S.,  {de Winter} D.,    {Perez} M.~R.,  1994, \aaps, 104, 315

\bibitem[\protect\citeauthoryear{{Tody}}{{Tody}}{1993}]{Tody93}
{Tody} D.,  1993, in {Hanisch} R.~J.,  {Brissenden} R.~J.~V.,   {Barnes} J.,
  eds, Astronomical Data Analysis Software and Systems II Vol.~52 of
  Astronomical Society of the Pacific Conference Series, {IRAF in the
  Nineties}.
p.~173

\bibitem[\protect\citeauthoryear{{Trammell}, {Dinerstein} \&
  {Goodrich}}{{Trammell} et~al.}{1994}]{trammell94}
{Trammell} S.~R.,  {Dinerstein} H.~L.,    {Goodrich} R.~W.,  1994, \aj, 108,
  984

\bibitem[\protect\citeauthoryear{{Vieira}, {Corradi}, {Alencar}, {Mendes},
  {Torres}, {Quast}, {Guimar{\~a}es} \& {da Silva}}{{Vieira}
  et~al.}{2003}]{2003AJ....126.2971V}
{Vieira} S.~L.~A.,  {Corradi} W.~J.~B.,  {Alencar} S.~H.~P.,  {Mendes}
  L.~T.~S.,  {Torres} C.~A.~O.,  {Quast} G.~R.,  {Guimar{\~a}es} M.~M.,    {da
  Silva} L.,  2003, \aj, 126, 2971

\bibitem[\protect\citeauthoryear{{Vink}}{{Vink}}{2015}]{vink15}
{Vink} J.~S.,  2015, \apss, 357, 98

\bibitem[\protect\citeauthoryear{{Vink}, {Drew}, {Harries} \&
  {Oudmaijer}}{{Vink} et~al.}{2002}]{Vink2002}
{Vink} J.~S.,  {Drew} J.~E.,  {Harries} T.~J.,    {Oudmaijer} R.~D.,  2002,
  \mnras, 337, 356

\bibitem[\protect\citeauthoryear{{Vink}, {Drew}, {Harries}, {Oudmaijer} \&
  {Unruh}}{{Vink} et~al.}{2005}]{vink05a}
{Vink} J.~S.,  {Drew} J.~E.,  {Harries} T.~J.,  {Oudmaijer} R.~D.,    {Unruh}
  Y.,  2005, \mnras, 359, 1049

\bibitem[\protect\citeauthoryear{{Vink}, {Drew}, {Harries}, {Oudmaijer} \&
  {Unruh}}{{Vink} et~al.}{2003}]{vink03}
{Vink} J.~S.,  {Drew} J.~E.,  {Harries} T.~J.,  {Oudmaijer} R.~D.,    {Unruh}
  Y.~C.,  2003, \aap, 406, 703

\bibitem[\protect\citeauthoryear{{Vink}, {Harries} \& {Drew}}{{Vink}
  et~al.}{2005}]{vink05b}
{Vink} J.~S.,  {Harries} T.~J.,    {Drew} J.~E.,  2005, \aap, 430, 213

\bibitem[\protect\citeauthoryear{{Vrba}, {Schmidt} \& {Hintzen}}{{Vrba}
  et~al.}{1979}]{1979ApJ...227..185V}
{Vrba} F.~J.,  {Schmidt} G.~D.,    {Hintzen} P.~M.,  1979, \apj, 227, 185

\bibitem[\protect\citeauthoryear{{Wheelwright}, {Bjorkman}, {Oudmaijer},
  {Carciofi}, {Bjorkman} \& {Porter}}{{Wheelwright}
  et~al.}{2012}]{wheelwright12}
{Wheelwright} H.~E.,  {Bjorkman} J.~E.,  {Oudmaijer} R.~D.,  {Carciofi} A.~C.,
  {Bjorkman} K.~S.,    {Porter} J.~M.,  2012, \mnras, 423, L11

\bibitem[\protect\citeauthoryear{{Wheelwright}, {Vink}, {Oudmaijer} \&
  {Drew}}{{Wheelwright} et~al.}{2011}]{2011A&A...532A..28W}
{Wheelwright} H.~E.,  {Vink} J.~S.,  {Oudmaijer} R.~D.,    {Drew} J.~E.,  2011,
  \aap, 532, A28

\bibitem[\protect\citeauthoryear{{Whittet}, {Martin}, {Hough}, {Rouse},
  {Bailey} \& {Axon}}{{Whittet} et~al.}{1992}]{Whittet92}
{Whittet} D.~C.~B.,  {Martin} P.~G.,  {Hough} J.~H.,  {Rouse} M.~F.,  {Bailey}
  J.~A.,    {Axon} D.~J.,  1992, \apj, 386, 562

\bibitem[\protect\citeauthoryear{{Yudin} \& {Evans}}{{Yudin} \&
  {Evans}}{1998}]{1998A&AS..131..401Y}
{Yudin} R.~V.,  {Evans} A.,  1998, \aaps, 131, 401

\end{thebibliography}
\appendix
\section[]{Observed polarisation in the optical range from $\sim$4600~$\rm\AA$ to $\sim$9400~${\rm \AA}$ for all objects}

\begin{landscape}
\begin{figure}
        \centering
        \includegraphics[width=17cm,height = 23cm,angle = 270]{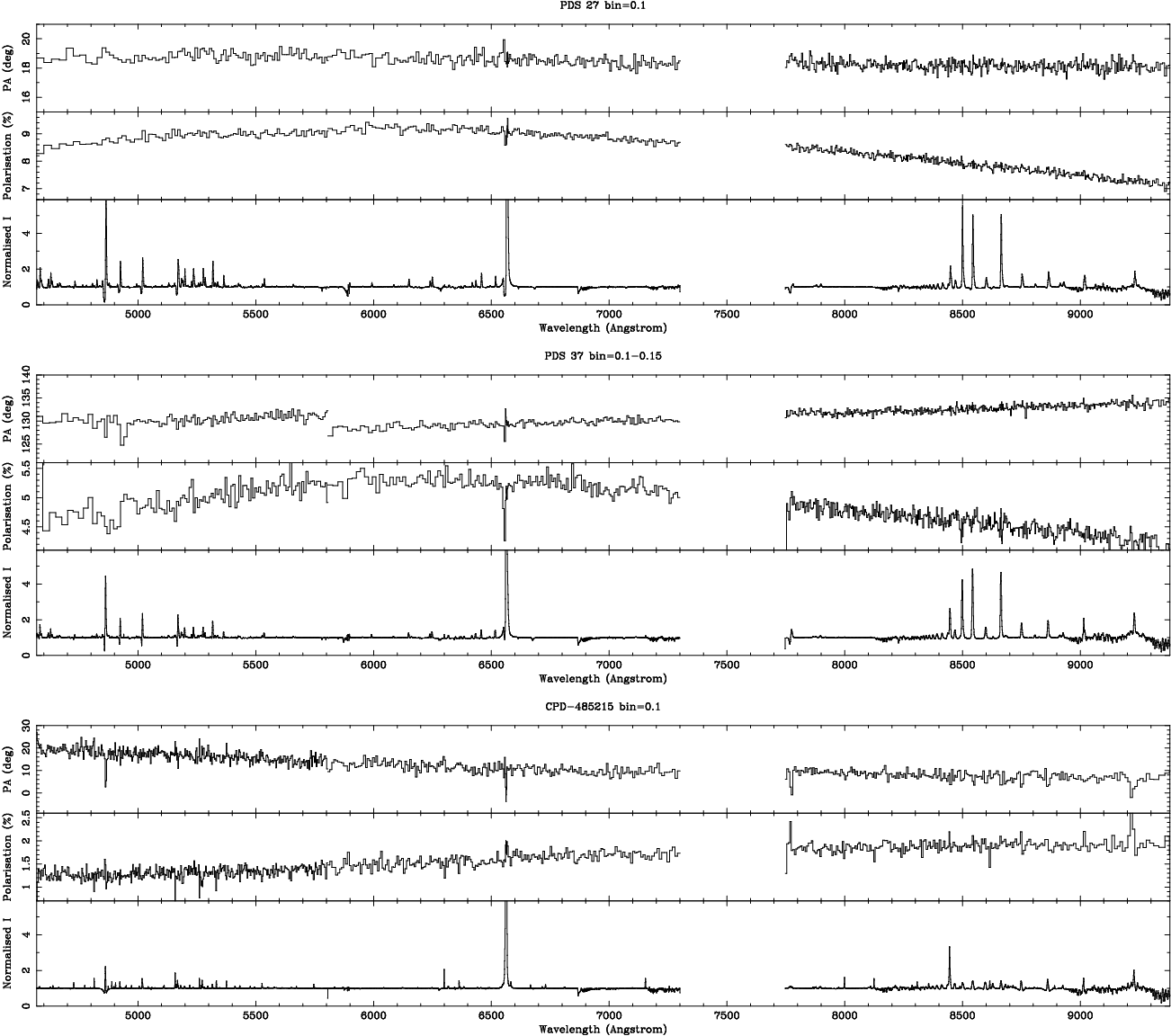}

   \caption{The polarisation data in {\it V}, {\it R} and {\it Z} bands for PDS 27, PDS 37 ,CPD-485215, R Mon and V380 Ori; and in {\it R} and {\it Z} bands for GU CMa, HD 98922, PDS 133, HD 104237, MWC 275, BF Ori and HD 85567.  R Mon, V380 Ori and HD 85567 are presented in two epochs in {\it R}. The data are presented as a triplots. In the
     triplot polarisation spectra the Stokes intensity (I) is shown in the lower
    panel, polarisation (\%) in the centre, while the
     position angle (PA) is shown in the upper panel. The data are rebinned to a constant value, as indicated at the top of each plot.}
  
\end{figure}
\end{landscape}

\begin{landscape}
\setcounter{figure}{0} 
\begin{figure}
        
        \centering
       \includegraphics[width=17cm,height = 23cm,angle = 270]{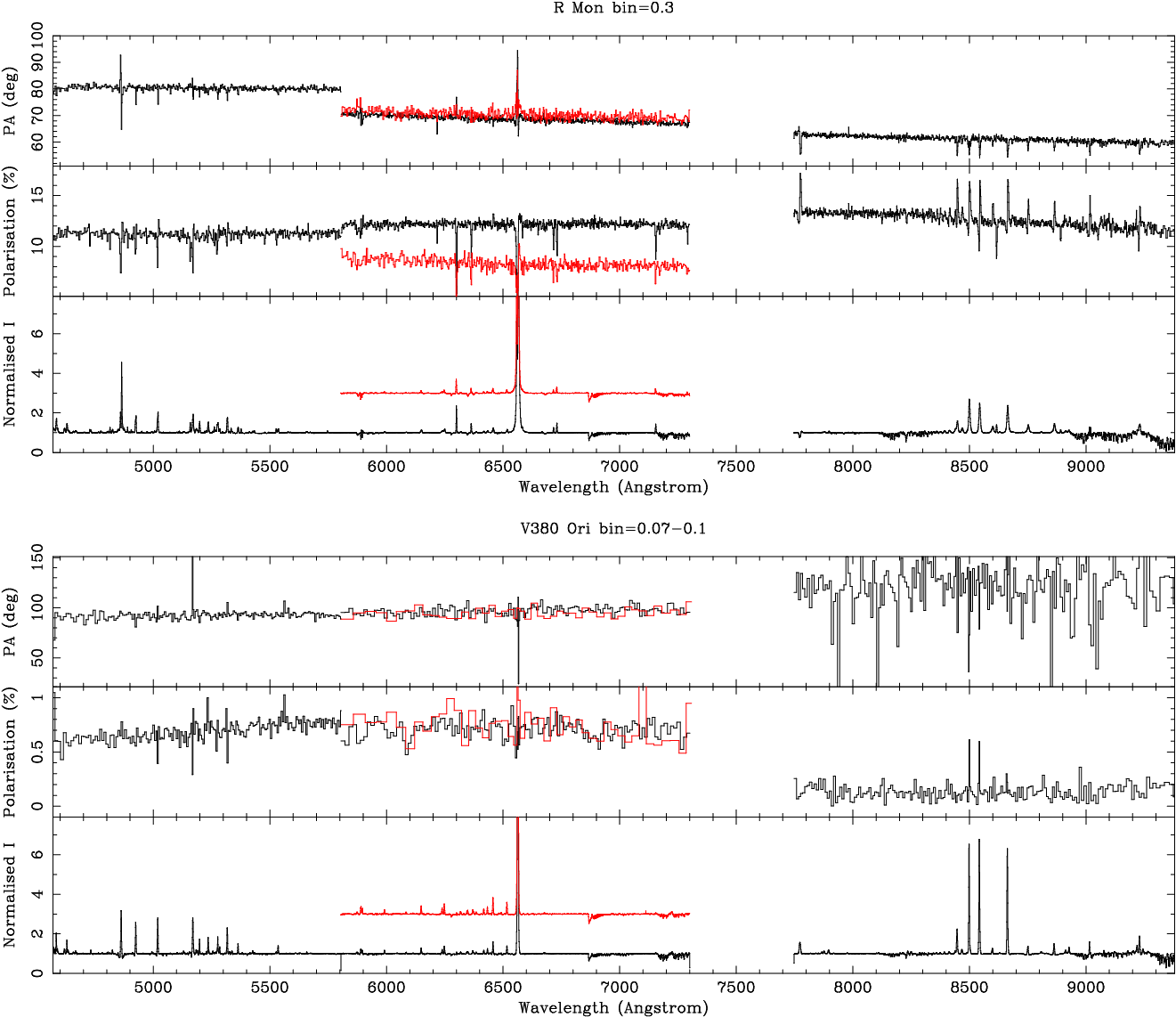}
       
   \caption{continued}
   
\end{figure}
\end{landscape}

\setcounter{figure}{0} 
\begin{figure*}
        
        \centering
       \includegraphics[width=17cm,height = 23cm,angle = 0]{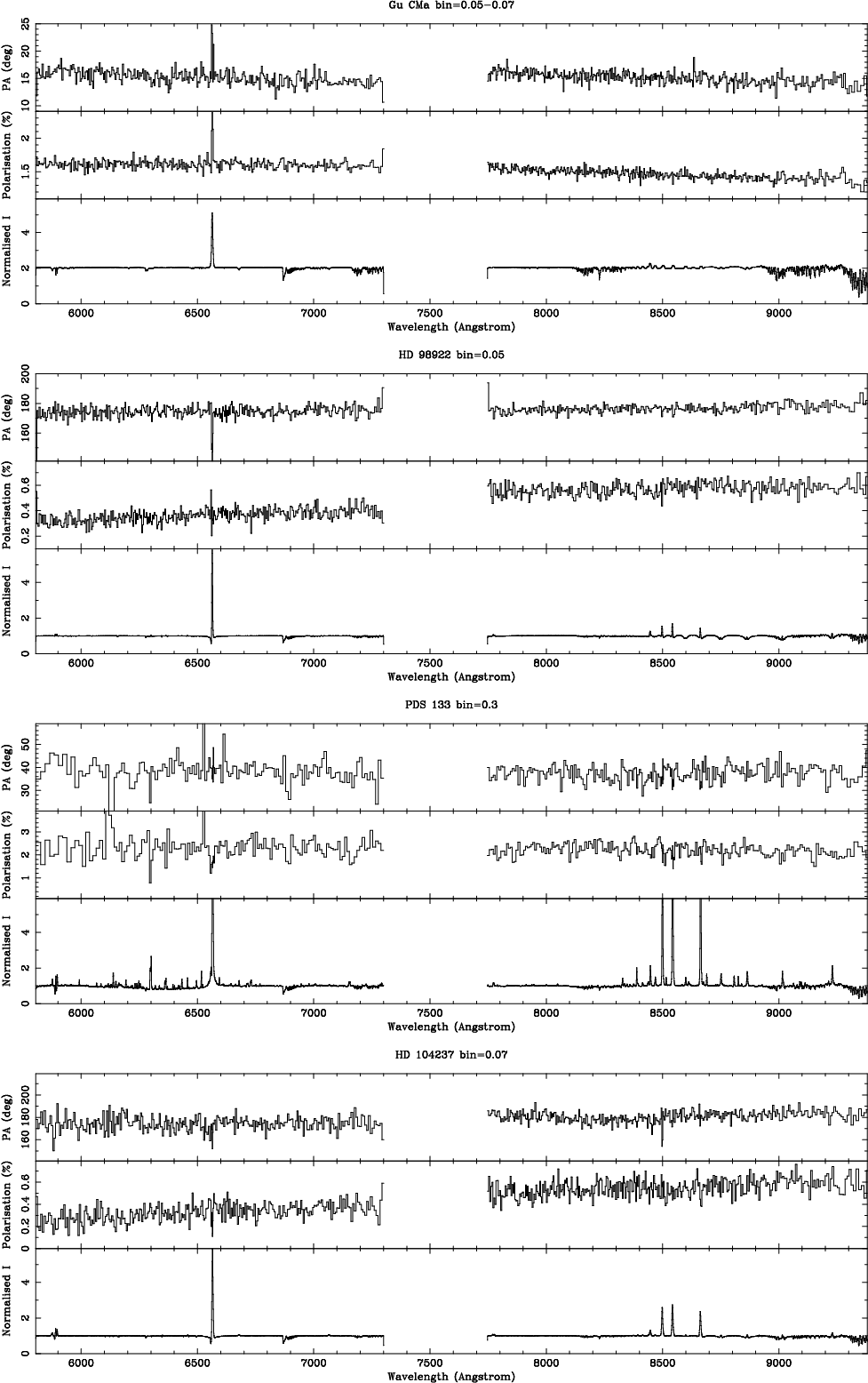}
       \label{a}

 \caption{continued}
   
\end{figure*}

\setcounter{figure}{0} 
\begin{figure*}
        
        \centering
       \includegraphics[width=17cm,height = 23cm,angle = 0]{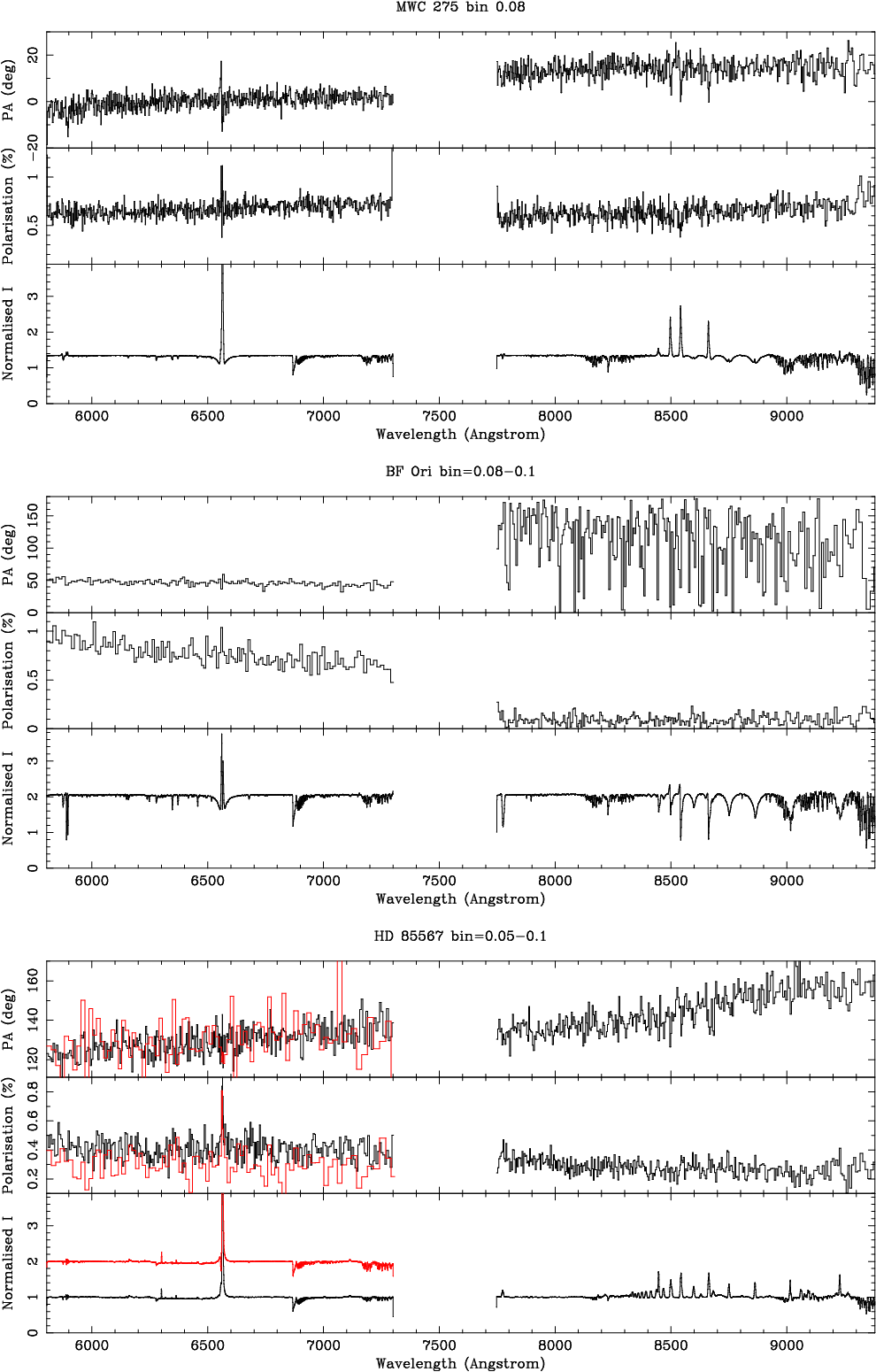}

 \caption{continued}
 \label{all}     
\end{figure*}

\label{lastpage}
\end{document}